\documentclass[10pt, conference, letterpaper]{IEEEtran}
\IEEEoverridecommandlockouts
\usepackage{cite}
\usepackage{amsmath,amssymb,amsfonts}
\usepackage{graphicx}
\usepackage{textcomp}
\usepackage{xcolor}
\usepackage{array}
\usepackage{rotating}
\usepackage{multirow}
\usepackage{amsthm}
\usepackage{float} 
\usepackage[font={footnotesize}]{subcaption}
\usepackage{booktabs}
\usepackage{url}
\usepackage{verbatim}

\usepackage[shortlabels]{enumitem}

\usepackage{mathrsfs}
\usepackage{xfrac}

\allowdisplaybreaks

\usepackage{thmtools}

\declaretheoremstyle[%
 spaceabove=4pt,%
 spacebelow=4pt,%
 headfont=\normalfont\bfseries,%
 bodyfont=\normalfont\itshape,%
 postheadspace=0.5em,%
]{theoremstyle} 

\declaretheorem[name={Definition},style=theoremstyle]{definition}
\declaretheorem[name={Assumption},style=theoremstyle]{assumption}
\declaretheorem[name={Theorem},style=theoremstyle]{theorem}
\declaretheorem[name={Corollary},style=theoremstyle]{corollary}

\declaretheorem[name={Proposition},style=theoremstyle]{proposition}

\declaretheoremstyle[%
 spaceabove=0pt,%
 spacebelow=4pt,%
 headfont=\normalfont\itshape,%
 postheadspace=1em,%
 qed=\qedsymbol%
]{proofstyle}

\def\BibTeX{{\rm B\kern-.05em{\sc i\kern-.025em b}\kern-.08em
    T\kern-.1667em\lower.7ex\hbox{E}\kern-.125emX}}
    
\newcommand{\Expect}{{\rm I\kern-.3em E}}
\newcommand{\Identity}{{\rm I\kern-.2em l}}

\usepackage[ruled,vlined,linesnumbered]{algorithm2e} 
\newcommand{\nosemic}{\renewcommand{\@endalgocfline}{\relax}}%
\newcommand{\dosemic}{\renewcommand{\@endalgocfline}{\algocf@endline}}%
\let\oldnl\nl%
\newcommand{\nonl}{\renewcommand{\nl}{\let\nl\oldnl}}%

\begin{document}

\title{Federated Learning While Providing Model as a Service: Joint Training and Inference Optimization\vspace{-0.2em}}

\author{
\IEEEauthorblockN{Pengchao Han\IEEEauthorrefmark{2}, Shiqiang Wang\IEEEauthorrefmark{3}, Yang Jiao\IEEEauthorrefmark{4}, Jianwei Huang\IEEEauthorrefmark{1}\IEEEauthorrefmark{5}}
\IEEEauthorblockA{\IEEEauthorrefmark{2}School of Information Engineering, Guangdong University of Technology, Guangzhou, China}
\IEEEauthorblockA{\IEEEauthorrefmark{3}IBM T. J. Watson Research Center, Yorktown Heights, NY, USA}
\IEEEauthorblockA{\IEEEauthorrefmark{4}Department of Computer Science and Technology, Tongji University, Shanghai, China}
\IEEEauthorblockA{\IEEEauthorrefmark{5}School of Science and Engineering, Shenzhen Institute of Artificial Intelligence and Robotics for Society, \\ The Chinese University of Hong Kong, Shenzhen, Shenzhen China}
 \IEEEauthorblockA{Emails: hanpengchao@gdut.edu.cn, shiqiang.wang@ieee.org, yangjiao@tongji.edu.cn,  jianweihuang@cuhk.edu.cn}
 \thanks{
 \IEEEauthorrefmark{1}Corresponding author.

 The work of P. Han and J. Huang was supported by the National Natural Science Foundation of China (Project 62271434), Shenzhen Science and Technology Innovation Program (Project JCYJ20210324120011032), Guangdong Basic and Applied Basic Research Foundation (Projects 2021B1515120008, 2022A1515110056), Shenzhen Key Lab of Crowd Intelligence Empowered Low-Carbon Energy Network (No. ZDSYS20220606100601002), and the Shenzhen Institute of Artificial Intelligence and Robotics for Society.}\vspace{-2em}
}

\maketitle

\begin{abstract}
While providing machine learning model as a service to process users' inference requests, online applications can periodically upgrade the model utilizing newly collected data. Federated learning (FL) is beneficial for enabling the training of models across distributed clients while keeping the data locally. However, existing work has overlooked the coexistence of model training and inference under clients' limited resources. This paper focuses on the joint optimization of model training and inference to maximize inference performance at clients. Such an optimization faces several challenges. The first challenge is to characterize the clients' inference performance when clients may partially participate in FL. To resolve this challenge, we introduce a new notion of age of model (AoM) to quantify client-side model freshness, based on which we use FL's global model convergence error as an approximate measure of inference performance. The second challenge is the tight coupling among clients' decisions, including participation probability in FL, model download probability, and service rates. Toward the challenges, we propose an online problem approximation to reduce the problem complexity and optimize the resources to balance the needs of model training and inference. Experimental results demonstrate that the proposed algorithm improves the average inference accuracy by up to 12\%.

\end{abstract}


\section{Introduction} 
\label{sec:intro}
Providing machine learning model as a service to support online applications, such as autonomous vehicles and industrial Internet of Things (IoT), often requires deploying models at the edge of the network. This enables efficient processing of user requests with low delays and also ensures the privacy of users' inference requests, enabling distributed model service provisioning (SP).
Frameworks like  TorchServe \cite{TorchServe} and TensorFlow Serving \cite{TensorFlowServing} facilitate the efficient and reliable deployment and management of models in production environments.
In essence, processing users' requests involves performing model inference on the request data. The quality of request processing is associated with the \textit{inference performance} provided to users and the \textit{service stability} reflected by finite processing delay.

\begin{figure}[t]
\centering
 \centering
 \includegraphics[width=0.45\textwidth]{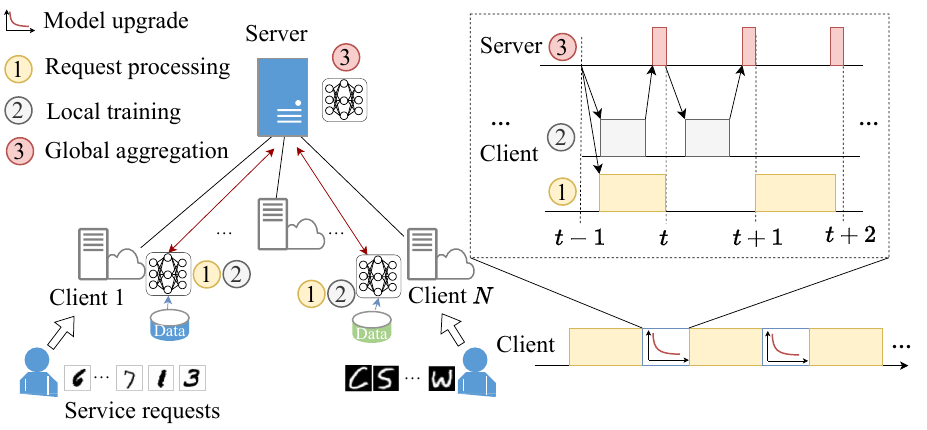}
\caption{\label{fig:architecture}  Joint FL and SP architecture.}
\vspace{-5pt}
\end{figure}

To enhance performance for online SP, the machine learning model should periodically be upgraded utilizing the newly collected user data. To this end, federated learning (FL) \cite{mcmahan2017communication, kairouz2019advances,sun2021pain,sun2022profit} enables the training of machine learning models across multiple distributed clients while keeping the data locally. 
Thus, during model upgrade, clients simultaneously perform FL to improve the model and inference to process users' requests, thereby preventing service interruption.
The joint FL and SP system architecture is shown in Fig. \ref{fig:architecture}. In each model upgrade process, the model is trained to converge over newly collected data using FL. During this process, the model training and request processing coexist on the same client. %

However, clients often have limited and heterogeneous computation and communication resource capabilities \cite{imteaj2021survey}, posing challenges for joint model training and inference with the following tradeoff.
\begin{itemize}
    \item Model training resource costs in FL: Efficient FL requires active clients' participation to accelerate the training process. The participating clients in each round of FL consume communication resources to download the global model from the server and send model updates to the server. Moreover, participating clients consume computation resources for local training.
    \item Model inference resource costs in SP: For SP, processing users' requests requires computation resources for performing model inference. Furthermore, since clients may not participate in all rounds of FL, a client with an outdated model generally achieves worse inference performance than those who have the latest global model. 
    To improve inference performance, a client can keep the model fresh by downloading the latest global model from the server (even if it does not participate in FL). Such downloading operation will also consume additional communication resources. 

\end{itemize}

Targeting maximizing inference performance at clients for SP while maintaining service stability, clients must balance resources for model training and inference in an adaptive manner. 
This motivates us to ask the following key question. 

\textbf{Key question}: \textit{How to jointly optimize model training and inference under resource constraints to maximize the overall inference performance?}

There are several challenges to answering this question.
\begin{itemize}
    \item \textbf{Challenge 1: Inference performance characterization.}
Characterizing the inference performance is critical for optimizing the performance of request processing for users. However, it is difficult because the model accuracy is non-differentiable, thus cannot be used as the objective directly.
Moreover, different clients may have different versions of the global model, due to clients' partial participation in FL.
The less fresh model will decrease the inference performance of the corresponding client. 

\item \textbf{Challenge 2: Non-convex objective over coupled decisions. }
The problem of maximizing the long-term inference performance is non-convex, which is challenging to solve. 
The problem is constrained by time-averaged resource budget constraints of clients. 
We cannot directly apply the traditional online decision-making algorithm to optimize the long-term objective with time-averaged constraints due to the unique coupling of clients' decisions in this problem. 
Specifically, the freshest model for request processing is obtained in the previous time slot from FL. 
This indicates that we cannot optimize the current inference performance by 
 optimizing the FL in the same time slot. 
\end{itemize}

To address the first challenge, we first characterize the global model performance by analyzing the convergence error upper bound of FL considering %
clients' partial participation in model training. We consider multiple local updates and non-convex loss functions in the bound. %
Then, we introduce the notion of \textit{age of model} (AoM) to capture each client's model freshness. 
The AoM measures the number of time slots by which a client’s local model falls behind the latest global model. A client can reduce its AoM by downloading the global model from the server, which can occur in both FL and SP.
Finally, we characterize the client's long-term inference performance by a surrogate involving the convergence error of the global model, the AoM of the client, and service rate. %

To address the second challenge, we first approximate the inference performance optimization problem with an online decision-making problem to asymptotically satisfy time-averaged resource constraints. The objective is biconvex with respect to each variable. Thus, we then apply alternating optimization of model training and inference. 
To address the challenge of coupled clients' decisions, we propose to decide clients' participation probabilities in FL for the current time slot and the download probabilities and the service rate for the subsequent time slot simultaneously.

The main contributions of this paper are as follows.
\begin{itemize}
     \item \textbf{Joint FL and SP framework.} To the best of our knowledge, we are the first to propose a joint FL and SP framework to maximize inference performance and ensure service stability while upgrading the model.  
     We propose an online federated learning and service provisioning (FedLS) algorithm to tackle the challenging problem of joint model training and inference optimization. 
     \item \textbf{Clients' inference performance characterization.} To address the challenge of inference performance characterization, we first characterize the global model performance using the convergence error upper bound of FL, specifically addressing the impact of partial participation in FL. 
     Then, we define the AoM to characterize the model freshness at clients, which measures the amount of time that the client's model falls behind the latest global model. 
     Finally, we approximate the inference performance at clients by considering global model convergence error and clients' AoM. %
     \item \textbf{Online algorithm design.} To balance resources for model training and inference while addressing the unique coupling of clients' decisions, we propose the FedLS algorithm.  
     By optimizing clients' decisions in each time slot, we satisfy the time-averaged computation and communication resource constraints with bounded violation error.
     By alternatively optimizing clients' participation probabilities in model training in each time slot and the download probabilities and service rates in request processing in the subsequent slot, we maximize the inference performance with service stability. 
     \item \textbf{Experiments on real-world datasets.} We conduct experiments on real-world datasets to validate the effectiveness of FedLS. The results demonstrate that FedLS achieves superior average inference accuracy for clients by up to 12\% with lower processing delay than the baseline. 
 \end{itemize}
\section{Related Work}\label{sec:related_work}

Researchers have proposed various approaches to address the challenge of resource-constrained federated learning (FL). These works mainly leverage clients' partial participation and sparse communication.
Partial participation in FL involves selecting a fraction of clients to participate in FL in each time slot, aiming to reduce the average computation costs of clients (e.g., \cite{Cho2022towards,zhou2023rein,ding2023incentive,jiang2023hetero,9484767,9435350,9302575,fraboni2021clustered,yang2021achieving,li2019convergence, 9053740,wang2022age,maaoi}). 
To further reduce the communication resource overhead, researchers have employed sparse communication techniques by optimizing the amount of communication between clients and the server through adjusting the number of local updates per communication round (e.g., \cite{zhou2023rein,wang2019adaptive,liao2023adap}), gradient compression (e.g., \cite{jiang2023hetero,8889996,gorbunov2021marina,alistarh2018convergence,stich2019error,hegde2023network,li2023anyc}), 
model pruning (e.g., \cite{9707474_dropout}),
and gradient quantization (e.g., \cite{Alistarh2017qsgd,reisizadeh2020fedpaq,liu2023commu,wang2022fedlite}).
\emph{However, these prior works focused on optimizing the model training performance, without considering online inference request processing of the trained models. }

The impact of the information freshness \cite{8514816} has been considered in the online optimization for machine learning tasks. In \cite{9484640}, the authors analyzed how the freshness of data affects the loss minimization in model training. In FL, the freshness of model update has been taken into account when selecting clients to participate in model training, with a focus on accelerating the overall training process ( e.g., \cite{9053740,wang2022age,maaoi}).
\emph{However, the impact of clients' model freshness on the inference performance of the clients has not been studied. }

To the best of our knowledge, there does not exist prior work that tackles the challenge of optimizing inference performance while incorporating both distributed model training and inference. 

\section{Joint FL and SP Optimization Problem}
In this section, we present the joint FL and SP system in Sec. \ref{sec:framework} and present clients' decisions in Sec. \ref{sec:client_decisions}. We introduce the resource capacities and cost and formulate the inference performance at clients as objective in Secs. \ref{sec:resource} and \ref{sec:qos_characterization}, respectively. Last, we formulate the joint  FL and SP optimization problem in Sec. \ref{sec:problem}.

\subsection{System Model}\label{sec:framework}
\textbf{System Overview.} The joint FL and SP system includes a set $\mathcal{N}=\{1, ..., N\}$ of clients. 
We consider an entire model upgrade period of joint FL and SP process consisting of $T$ time slots (as shown in Fig. \ref{fig:architecture}).
Each client~$n$ has a local dataset $\mathcal{D}_n$, which remains unchanged during this period. 
The inference tasks stochastically arrive at each client during this period. The clients need to upgrade the model (through FL on the distributed local datasets at clients) and process inference requests in parallel during this period. After this period, the clients will continuously obtain some additional new data, based on which a new period may start at a later time. In this paper, we focus on a single period where FL and SP coexist.

\textbf{FL in Slot $t$.} 
In FL, each client $n$ trains a local model using its local dataset to minimize loss function $F_n\left(\mathbf{x}\right)$. 
All clients collaborate to train a global model $\mathbf{x}$ to solve the following global loss minimization problem:
\begin{align}
    \min_{\mathbf{x}} \quad f\left(\mathbf{x}\right):=\frac{1}{N}\sum\nolimits_{n=1}^NF_n\left(\mathbf{x}\right).
\end{align}
In each time slot, the clients perform one round of local model training and global aggregation.  Note that only part of the clients may participate in the training in a particular time slot, and the participating client set may change over time. 

\textbf{SP in Slot $t$.} For SP, we denote the arrival rate of inference requests for client $n$ at time slot $t$ by $\lambda_t^n$. Each client maintains a service queue, within which the inference requests are buffered and served in a first-in-first-out manner.

\textbf{Overall Working Flow.} Each time slot $t$ involves three main steps of joint FL and SP, as shown in Fig. \ref{fig:architecture}.
\begin{itemize}
    \item Parallel local model training and request processing
    \begin{itemize}
        \item \textbf{Inference request processing for SP:} %
        Using its local model, each client processes a set of inference requests in the service queue.
        \item \textbf{Model training for FL:} %
        Each participating client trains its local model using its local dataset, starting from $\mathbf{x}_{t}$  for $\tau$ iterations.
    \end{itemize}
    \item \textbf{Global model aggregation for FL:} %
    The server aggregates the local model parameters from all participating clients to obtain an updated global model $\mathbf{x}_{t+1}$. 
\end{itemize}

\subsection{Clients' Decisions} \label{sec:client_decisions} 

\textbf{Clients' Decisions in Model Training.} 
In each time slot $t$, each client $n$ determines its participation probability in FL, denoted by $q^n_t \in \left(0,1\right]$. Based on the probability, each client then decides whether to participate in FL or not, represented by $\mathbf{I}_t^n\in \left\{0,1\right\}$, by sampling from the Bernoulli distribution with parameter $q_t^n$. If a client participates in model training, it needs to synchronize its local model with the latest global model by downloading the global parameter from the server. This ensures that all participating clients start with the same latest version of the global model.  We denote $\boldsymbol{q}=\left\{q^n_t,\forall n,t\right\}$.

\textbf{Clients' Decisions in Request Processing.} 
For request processing, a client needs to determine the download probabilities $\beta_t^n,\forall t$ and service rates $\mu_t^n, \forall t$, where $\beta_t^n$ indicates the probability of client $n$ downloading the global model from the server for both model training and request processing in time slot $t$.
The rationale behind $\beta_t^n$ is that clients with an outdated local model can 
download the latest global model from the server to process inference requests with high performance. 
We denote $\mathbf{J}_t^{n}\in\left\{0,1\right\}$ as the download decision for request processing sampled from the Bernoulli distribution with parameter $\beta_t^n$.
Meanwhile, if $\mathbf{J}_t^{n}=1$, the client does not need to download the global model for model training in the same time slot, as the client already has the latest global model.This implies that $\mathbf{I}_t^{n}$ and $\mathbf{J}_t^{n}$ are not independent. In particular, if $\mathbf{I}_t^{n}=1$, then we always have $\mathbf{J}_t^{n}=1$. 
Furthermore, to ensure service stability while satisfying resource constraints, client $n$ needs to determine a service rate in each time slot $t$, denoted by $\mu_t^n$. %
For convenience, we denote $\boldsymbol{\beta}=\left\{\beta^n_t,\forall n,t\right\}$ and $\boldsymbol{\mu}=\left\{\mu^n_t,\forall n,t\right\}$.

\subsection{Resource Capacities and Costs} \label{sec:resource}
\textbf{Resource Capacities.} Each client has limited computation and communication resource capacities. 
We denote the computation capacity of client $n$ as $\tilde{\phi}^n$, and the communication capacity as $\tilde{\psi}^n$. Moreover, we use $\tilde{\phi}_{\max}^n$ and $\tilde{\psi}_{\max}^n$ to represent the maximum instantaneous computation and communication costs, respectively, that each client $n$ can consume in each time slot, reflecting its highest instant processing capabilities.

\textbf{Computation Cost.} The computation cost of a client depends on the participation probability in FL and the service rate of the client. 
We denote $\phi_t^n\left(q_t^n, \mu_t^n\right)$ as the computation cost of client $n$ in time slot $t$, which depends on $q_t^n$ and $\mu_t^n$. 

\textbf{Communication Cost.} We consider the downstream communication cost from the server to clients\footnote{It is worth noting that while we currently focus on the downstream communication costs of clients, our problem formulation and proposed algorithm can be easily extended to incorporate upstream communication costs as well.}. 
A client may download the global model from the server to participate in FL or to process inference requests.
Thus, we denote the communication cost of client $n$ in time slot $t$ by $\psi_t^n\left(q_t^n,\beta_t^n\right)$, which depends on $q_t^n$ and $\beta_t^n$.

\subsection{Inference Performance Characterization} \label{sec:qos_characterization}
We first formulate the AoM at clients considering the download operation of clients in model training and inference. Then, we characterize the inference performance using AoM and global model performance.

\textbf{AoM.} We introduce the notion of AoM to capture the model freshness at clients. Let $A_t^n$ be the AoM of client $n$ in time slot $t$. We have
\begin{definition}
    The AoM of a client is the number of time slots by which a client's local model falls behind the latest global model on the server.
\end{definition}

\begin{figure}[t]
\centering
 \centering
 \includegraphics[width=0.44\textwidth]{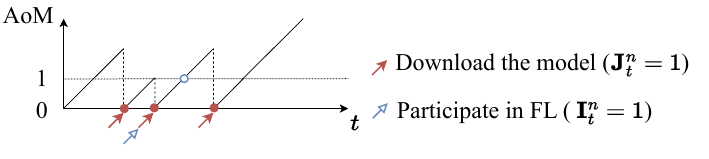}
\caption{\label{fig:aom}  Illustration of AoM.} %
\end{figure}

Formulating the AoM of a client is challenging because the freshness of the model on the client side is influenced by the client's download decisions in both model training and request processing. Let $G_t\left(\boldsymbol{q}\right)$ be the performance of the global model obtained at the start of time $t$, there are three possible values of $A_t^n$ based on $A_{t-1}^n$ as illustrated in Fig. \ref{fig:aom}:
\begin{itemize}
    \item If the client $n$ downloads the global model (obtained at the end of the previous time slot) from the server in time slot $t$, i.e., $\mathbf{J}_t^{n}=1$, the client processes requests with performance $G_{t}\left(\boldsymbol{q}\right)$. Thus, the AoM of the client is 0.
    \item If $\mathbf{J}_t^{n}\neq 1$, but the client participates in FL in the previous time slot, i.e., $\mathbf{I}_{t-1}^n=1$, the client processes requests with performance $G_{t-1}\left(\boldsymbol{q}\right)$, resulting an AoM of 1.
    \item Otherwise, if the client $n$ neither downloads the global model for model training in the previous time slot nor for request processing in the current time slot, its AoM increases by 1.
\end{itemize}

Overall, the probabilities of the AoM taking different values for client $n$ in time slot $t$ are as follows:
\begin{equation}\label{eq:aom_evol}
A_t^n\!=\!\left\{
\begin{aligned}
&0, {\rm with\ probability\ }\beta_t^n, \\
&1, {\rm with\ probability\ }\left(1-\beta_t^n\right)q_{t-1}^n,\\
&A_{t-1}^n\!+\!1, {\rm with\ probability\ }\left(1\!-\!\beta_t^n\right)\!\left(1\!-\!q_{t-1}^n\right),
\end{aligned}
\right.
\end{equation}
where we separately list the last two cases involving $q_{t-1}^n$ to facilitate the control decisions in Section~\ref{sec:alg-design}.
We can see from \eqref{eq:aom_evol} that the AoM of client $n$ in slot $t$ can be any value from 0 to $t$.
When $A_t^n=t'$, it means that the client downloaded the global model at $t\!-\!t'$ and did not download the model from $t\!-\!t'$ to the current time $t$.%

\textbf{Inference Performance.} 
The inference performance at client $n$ and time $t$ is the performance, e.g., accuracy, of the model that the client uses for inference request processing at that time. %
Based on the AoM of a client, the expected model performance that client $n$ provides at time $t$, denoted by $K_t^n(\boldsymbol{q},\boldsymbol{\beta})$, is
\begin{equation}
    K_t^n(\boldsymbol{q},\boldsymbol{\beta}) :=\sum\nolimits_{a=0}^{t}G_{t-a} \left(\boldsymbol{q}\right)P\left[A_t^n=a\right].
\end{equation}%
We characterize the inference performance using $K_t^n(\boldsymbol{q},\boldsymbol{\beta})$. 
In the later Section \ref{sec:convergence}, we will formulate $K_t^n(\boldsymbol{q},\boldsymbol{\beta})$ using the global model convergence error bound. 
Then, we maximize the inference performance at clients by minimizing $K_t^n(\boldsymbol{q},\boldsymbol{\beta}), \forall n$, as the model accuracy is non-differentiable and cannot be used as the objective directly.

\subsection{Problem Formulation} \label{sec:problem}

We formulate the joint FL and SP optimization problem as the following Problem $\mathbf{P1}$.  Our problem formulation allows for extensions to incorporate various resource budgets, such as energy consumption, enabling comprehensive resource management in joint FL and SP system.
    \begin{subequations}
     \begin{align}
     \mathbf{P1}: \min_{\boldsymbol{q},\boldsymbol{\beta},\boldsymbol{\mu}} \, %
     &{\!M_{T}}(\boldsymbol{q},\boldsymbol{\beta},\boldsymbol{\mu}) \!:=\!\sum\nolimits_{t=0}^{T-1}\!\sum\nolimits_{n=1}^N\!\mu_t^n K_t^n(\boldsymbol{q},\boldsymbol{\beta}) .\tag{4}\label{eq:p1_obj}\\
   {\rm s.t.} \quad & \textstyle\frac{1}{T}\sum\nolimits_{t=0}^{T-1} \mathbb{E}\left[\lambda_t^n-\mu_t^n\right] \leq 0, \forall n, \label{eq:cnst_service}\\
       & \textstyle\frac{1}{T}\sum\nolimits_{t=0}^{T-1} \mathbb{E}\left[\phi_t^n\left(q_t^n, \mu_t^n\right)\right] \leq \tilde{\phi}^n, \forall n,\label{eq:cnst_comp} \\
        & \textstyle\frac{1}{T}\sum\nolimits_{t=0}^{T-1} \mathbb{E}\left[\psi_t^n\left(q_t^n, \beta_t^n\right)\right] \leq \tilde{\psi}^n, \forall n,\label{eq:cnst_commu}\\
        & \phi_t^n\left(q_t^n, \mu_t^n\right) \leq \tilde{\phi}_{\max}^n,\forall n,t,\label{eq:cnst_comp1}\\
        & \psi_t^n\left(q_t^n, \beta_t^n\right) \leq \tilde{\psi}_{\max}^n,\forall n,t,\label{eq:cnst_comm1}\\
         \quad &  q_t^n \in \left(0,1\right], \beta^n_t\in \left[0,1\right], \mu_t^n \geq 0, \forall  n, t. \label{eq:cnst_variable}
    \end{align}
    \end{subequations}
    
Problem $\mathbf{P1}$ aims to maximize the long-term inference performance at all clients over $T$ time slots, denoted by $M_{T}(\boldsymbol{q},\boldsymbol{\beta},\boldsymbol{\mu})$. 
Constraint \eqref{eq:cnst_service} ensures mean rate stability of service queues for clients, preventing infinite queue growth. Constraints \eqref{eq:cnst_comp} and \eqref{eq:cnst_commu} impose time-averaged computation and communication resource budgets on each client, limiting average resource consumption.
Additionally, Constraints \eqref{eq:cnst_comp1} and \eqref{eq:cnst_comm1} restrict the instantaneous computation and communication costs of each client in each time slot.
We consider the participation probability of client $n$ for model training, i.e., $q_t^n$, to be constrained within $\left(0,1\right]$. The download probability $\beta_t^n$ is within $\left[0,1\right]$ and the service rate $\mu_t^n$ is a non-negative value.
When it is clear from the context, we omit the variables $q_t^n$, $\beta_t^n$ and $\mu_t^n$ and write the costs as $\phi_t^n$ and $\psi_t^n$, the model performance as $G_t$ and $K_t^n$, and the objective as $M_t$ in short. 

\textbf{Challenges.}
Solving Problem $\mathbf{P1}$ poses significant challenges. 
First, the impact of clients' partial participation $\boldsymbol{q}$ in FL on global model performance $G_t$ is hard to formulate.
Second, Problem $\mathbf{P1}$ includes long-term objective \eqref{eq:p1_obj} and time-averaged constraints \eqref{eq:cnst_service}--\eqref{eq:cnst_commu}. 
Decisions in history would affect future decisions. The unpredictable future resource costs complicate finding optimal solutions in advance.
Third, clients' decisions are coupled across time slots. The inference performance in slot $t$, i.e., $M_t$, depends on the model performance obtained in the previous slot, i.e., $G_{t-1}$, but not $G_t$.

To address these challenges, we use the theoretical convergence error upper bound considering clients' partial participation in FL to characterize the global model performance. The approach has been widely used in the FL literature \cite{wang2022federated,luo2021cost}. 
Then, we propose an online decision-making approach based on the Lyapunov drift-plus-penalty framework~\cite{bookneely} to handle long-term objective and correlated constraints effectively. 
In addition, in response to the challenge of coupled clients' decisions, we perform alternating optimization \cite{gorski2007biconvex, 9705079} for model training and inference in the adjacent time slots, while respecting the shared resource cost constraints.

\section{Problem Approximation and Algorithm Design}
\label{sec:alg-design}
We first approximate Problem $\mathbf{P1}$ using an online decision-making problem in Sec. \ref{sec:online-decision-making}.
Then, we solve the problem using alternating optimization and propose the online FedLS algorithm in Sec. \ref{sec:alternative-opt}. We analyze the time-averaged constraint satisfaction by using the proposed online algorithm in Sec. \ref{sec:constraint_satis}. Finally, we present the closed-form solutions considering linear resource costs in Sec. \ref{sec:close-form}.

\subsection{Online Decision Making} \label{sec:online-decision-making}
We approximate Problem $\mathbf{P1}$ using an online decision-making framework to tackle the challenges posed by long-term objective and time-averaged constraints.
First, to address the long-term objective in \eqref{eq:p1_obj}, we formulate the evolution of long-term inference performance at clients across time slots. 
Then, we define virtual queues to transform the time-averaged resource constraints into queue stability problems.
Finally, we present the approximate problem in each time slot.

\subsubsection{{Evolution of Objective}} \label{sec:convergence}
To analyze the change of the objective \eqref{eq:p1_obj} in each time slot, we first approximate  $G_t$ using the convergence error upper bound.  Then, we derive the evolution of $G_t$ and $M_t$ across time. 

\textbf{Convergence Error of FL.}
We analyze the convergence bound of FL with partial participation and multiple local updates represented by $\tau$. Let $\mathbf{g}_n\left(\cdot\right)$ be the stochastic gradient of client $n$, we make the following assumptions that are commonly used in the literature~\cite{wang2022federated,pmlr-v119-karimireddy20a,50448,jhunjhunwala2023fedexp,yu2019parallel,perazzone2022communication}.
\begin{assumption} \label{asm:overall} For each client $n$ and model parameters $\mathbf{x}$ and $\mathbf{y}$, we have
\begin{itemize}
    \item (Lipschitz gradient) There exists a constant $L\!>\!0$ that satisfies $\left\Vert\nabla F_n\left(\mathbf{x}\right)-\nabla F_n\left(\mathbf{y}\right)\right\Vert\leq L\left\Vert\mathbf{x}-\mathbf{y}\right\Vert$.
\item (Unbiased stochastic gradient with bounded variance) There exists a constant $\sigma>0$, such that  $\mathbb{E}[\mathbf{g}_n\left(\mathbf{x}\right)\big| \mathbf{x} ] = \nabla F_n\left(\mathbf{x}\right)$ and $\mathbb{E}[\left\Vert \mathbf{g}_n\left(\mathbf{x}\right)-\nabla F_n\left(\mathbf{x}\right) \right\Vert^2\big| \mathbf{x} ]\le\sigma^2$.

\item (Bounded gradient divergence) There exists a constant $\epsilon>0$, such that  $\left\Vert \nabla F_n\left(\mathbf{x}\right)-\nabla f\left(\mathbf{x}\right)  \right\Vert^2 \le \epsilon^2$.
\end{itemize}
\end{assumption}
Here, $\epsilon$ captures the data heterogeneity among clients. Note that we do not assume the convexity of clients' loss functions. For the first time, we simultaneously incorporate multiple local updates, clients' partial participation, and non-convex loss function into the convergence error bound of FL. Based on Assumption \ref{asm:overall}, we have the following theorem. 
\begin{theorem} \label{thm:convergence}
         Under Assumption \ref{asm:overall}, if the learning rate\footnote{Here, we consider a fixed learning rate over time. Nevertheless, it is worth noting that the analysis in this paper can be easily applied to a time-decay learning rate.}  $\eta\leq\frac{1}{\max\{L\tau,4L\tau \sum_{n=1}^N q_t^n/N\}}$, the convergence error of FL with partial participation and multiple local updates with $\tau$ iterations is upper bounded by
\begin{align}
    &\frac{1}{T}\sum\nolimits_{t=0}^{T-1}\mathbb{E}\left[\left\Vert\nabla f\left(\mathbf{x}_{t}\right)\right\Vert^2\right]\nonumber\\
    &\leq G_{T}\!\left(\boldsymbol{q}\right)\!:=\!\frac{4\!\left(f\left(\mathbf{x}_{0}\right)\!-\!f^\ast\right)}{T\tau\eta}\!+\!\frac{C}{TN}\!\sum\nolimits_{t=0}^{T-1}\!\sum\nolimits_{n=1}^N\!1/q_t^n,  \label{eq:conv1}
\end{align}  
where $C:=6L \eta \tau \left(\sigma^2+2\epsilon^2\right)$ and $f^{\ast}$ is the true minimum of $f\left(\mathbf{x}\right)$, i.e., $f^{\ast}:=\min_{\mathbf{x}}f\left(\mathbf{x}\right)$.
\end{theorem}
\textbf{Proof sketch:} The proof of Theorem~\ref{thm:convergence} is motivated by \cite{perazzone2022communication,wang2022federated} for considering clients' partial participation and \cite{50448} for considering multiple local updates in each communication round (time slot). In essence, let $\mathbf{x}_{t,i}^n$ be the local model parameter of client $n$ in $i$th iteration of round $t$. 
We first bound the local model parameter change in local training for each client in each time slot (i.e., $\sum_{i=0}^{\tau-1}  \mathbb{E}_t [ \left\Vert  \mathbf{x}_{t,i}^n- \mathbf{x}_{t}\right\Vert^2] $) based on Assumption~\ref{asm:overall} and Lemma 7 in \cite{50448}. 
Then, we bound the change of global model parameter in each time slot, (i.e., $\mathbb{E}_t \left[f\left(\mathbf{x}_{t+1}\right)\right]-f\left(\mathbf{x}_{t}\right)$), based on the model parameter update of $\mathbf{x}_{t+1} := \mathbf{x}_{t} - \frac{\eta}{N}\sum_{n=1}^N \sum_{i=0}^{\tau-1} \frac{\mathbf{I}_t^n}{q_t^n}\mathbf{g}_n\left(\mathbf{x}_{t,i}^n\right)$ for partial participation in FL \cite{wang2022federated}. %
Finally, by summing up global model parameter changes for $T$ time slots, we can obtain \eqref{eq:conv1}. \qed

\textbf{Evolution of Global Model Performance.} To characterize the change of global model performance $G_t$ in each time slot, we have the following evolution of $G_t$:
\begin{corollary} \label{col:convergence_round}
    Under Assumption \ref{asm:overall},
    the model convergence upper bound evolves as 
    \begin{align}\label{eq:evl_gt}
    G_{t}= \frac{t-1}{t}G_{t-1}+\frac{C}{Nt}\sum_{n=1}^{N}\frac{1}{q_{t-1}^n}.
    \end{align}
\end{corollary}
Corollary \ref{col:convergence_round} can be proved based on \eqref{eq:conv1}. The corollary reveals that the convergence error at time $t$ depends on both the convergence error at the previous time slot and the clients' participation probabilities in the current time slot. This empowers us to optimize the convergence of FL in an online fashion by dynamically adjusting the clients' participation probabilities for each time slot based on current performance. %

\textbf{Evolution of Clients' Inference Performance.}
Using \eqref{eq:aom_evol}, we can formulate the evolution of objective function in Problem $\mathbf{P1}$ from time $t$  to $t+1$ as follows.
\begin{align}
    M_{t+1} =M_{t}+\sum\nolimits_{n=1}^N&\mu_{t}^n\left( G_{t}\beta_{t}^n+G_{t-1}\left(1-\beta_{t}^n\right)q_{t-1}^n\right.\nonumber\\
    &\left.+K_{t-1}^n\left(1-\beta_{t}^n\right)\left(1-q_{t-1}^n\right)\right). \label{eq:evolution_mt}
\end{align}

From \eqref{eq:evolution_mt}, we can observe that the newly accumulated inference performance at clients in time slot $t+1$ is influenced by the possible model performance on the client side and the service rate in this time slot. The model performance can take values $G_t$, $G_{t-1}$, and $K_{t-1}^n$ with corresponding probabilities based on client~$n$'s download behaviors.  This observation motivates us to simultaneously optimize the model performance $G_t$, the download probability $\beta_{t}^n$, and service rate $\mu_{t}^n$ in each time slot $t$ based on known $G_{t-1}$ and $K_{t-1}^n$. Optimizing $G_t$ can be achieved by optimizing $q_{t-1}^n$ as indicated by \eqref{eq:evl_gt}. %

\subsubsection{{Service Queue and Virtual Queues}}
We transform the mean rate stable constraint \eqref{eq:cnst_service} and the time-averaged resource constraints \eqref{eq:cnst_comp} and \eqref{eq:cnst_commu} into queue stability problems to better manage resource allocation while ensuring constraint satisfaction. We formulate the evolution of queue lengths for these constraints as follows.

\textbf{Service Queue.} For the request processing of each client $n$, the queue length at time $t$, denoted by $\Lambda_t^n$, evolves as
\begin{align}
    & \Lambda_{t+1}^n=\max \left\{0, \Lambda_t^n+\lambda_t^n-\mu_t^n\right\}, \quad \forall n. \label{eq:queue_update}
\end{align}
Equation \eqref{eq:queue_update} captures the departures and arrivals of inference requests on clients in each time slot. %

\textbf{Virtual Queues.} We further define virtual queues to capture the constraint violation of \eqref{eq:cnst_comp}--\eqref{eq:cnst_commu}. 
We denote the queue lengths representing computation and communication costs by $\Phi_{t}^n$ and $\Psi_{t}^n$ respectively. We have
\begin{align}
& \Phi_{t+1}^n=\max \left\{0, \Phi_t^n+\phi_t^n-\tilde{\phi}_n\right\}, \quad \forall n. \label{eq:comp_update}\\
& \Psi_{t+1}^n=\max \left\{0, \Psi_t+\psi_t^n-\tilde{\psi}\right\}, \quad \forall n. \label{eq:comm_update}
\end{align}

The virtual queues effectively capture the consumed resources and the average available resources in each time slot. By ensuring the stability of these virtual queues, we can guarantee the satisfaction of the time-averaged resource constraints \eqref{eq:cnst_comp}--\eqref{eq:cnst_commu} with bounded violation error. This means that while there might be occasional resource violations in individual time slots, the average resource consumption over time will adhere to the specified constraints while time goes to infinite. 
Overall, the queues $\Lambda_t^n$, $\Phi_{t}^n$, and $\Psi_{t}^n$ capture the accumulated violation of constraints \eqref{eq:cnst_service}--\eqref{eq:cnst_commu}. 
We aim to jointly optimize the original objective in Problem $\mathbf{P1}$ and the queue stability of $\Lambda_t^n$, $\Phi_{t}^n$, and $\Psi_{t}^n$.

\subsubsection{{Decision Problem in Time Slot $t$}}
We approximate Problem $\mathbf{P1}$ using the Lyapunov drift-plus-penalty framework considering the original objective and the queue stability.  In Problem $\mathbf{P1}$, decisions are not independent across time slots. However, applying the Lyapunov drift-plus-penalty to approximately solve our problem is effective as motivated by \cite{wang2022federated, 8514816}.
Define a constant $V>0$, representing the importance of the objective compared to queues stability. 
Equation \eqref{eq:evolution_mt} motivates us to optimize clients' inference performance in a distributed manner. 
For each client in each time slot, we optimize the global model performance $G_{t+1}$ at time $t$ and the inference performance $M_{t+2}$ at the subsequent time slot $t+1$ simultaneously. The rationale behind this is that the best model performance that a client can provide in time slot $t$ is from the global model trained at the previous time slot. Therefore, in each time $t$, we optimize the clients' participation probabilities for the current time slot i.e., $q_t^n, \forall n$, and determine the model download probabilities and service rates for the subsequent time slot, i.e., $\beta_{t+1}^n, \mu_{t+1}^n, \forall n$, to obtain the optimal inference performance at time $t+1$.

Using Corollary~\ref{col:convergence_round} and \eqref{eq:evolution_mt}--\eqref{eq:comm_update}, we can approximate Problem $\mathbf{P1}$ using the Lyapunov drift-plus-penalty framework into the following Problem $\mathbf{P2-}t$ for each time $t$ to determine $q_t^n$, $\beta_{t+1}^n$, and $\mu_{t+1}^n$.
\begin{align}
    \textbf{P2-}{t}:&\min_{q_t^n, \beta_{t+1}^n,\mu_{t+1}^n} \!V\!\mu_{t\!+\!1}^n\!\left(\!\left(\!\frac{{t}}{{t\!+\!1}}\!G_{t}\!+\!\frac{C\!\sum_{n=1}^N 1/q_{t}^n}{N(t\!+\!1)}\right)\!\beta^n_{t+1}\!\right.\nonumber\\
    & \left.+G_{t}\left(1-\beta_{t+1}^n\right)q_{t}^n+K_{t}^n\left(1-\beta_{t+1}^n\right)\left(1-q_{t}^n\right)\right)\nonumber\\
    &+\!\Lambda_t^n\!\left(\!\lambda_t^n\!-\!\mu_t^n\!\right)\!+\!\Phi_t^n\!\left(\phi_t^n\!-\!\tilde{\phi}_t^n\right)\!+\!\Psi_t^n\!\left(\psi_t^n\!-\!\tilde{\psi}_t^n\right). \label{eq:obj_p2}\\
   &{\rm s.t.} \quad \eqref{eq:cnst_comp1}  -\eqref{eq:cnst_variable}\nonumber
\end{align}
By default, we set $\beta_0^n=0, \mu_0^n=0, \forall n$ to avoid processing requests with low accuracy because the model is randomly initialized with bad performance. %

In Problem $\mathbf{P2-}t$, our objective is to jointly optimize the inference performance at clients and (virtual) queue stability, balanced by the importance coefficient $V$, for the current time slot $t$.
In each time slot, we solve Problem $\mathbf{P2-}t$ conditioned on the real-time states of the system, including the local model performance $G^n_{t}, \forall n$, the global model performance $G_{t}$, and the queue lengths $\Lambda_{t}^n$, $\Phi_{t}^n$, and $\Psi_{t}^n$.

\subsection{Alternating Optimization and Algorithm Design } \label{sec:alternative-opt}
Directly solving Problem $\mathbf{P2-}t$ is difficult because the objective is non-convex. 
Moreover, we can observe from \eqref{eq:obj_p2} that if the client $n$ does not download the global model from the server at the subsequent time slot, i.e., $\beta_{t+1}^n=0$, the global model performance $G_t$ will not influence the objective $M_{t+2}$ in \eqref{eq:obj_p2}. However, improving $G_t$ is necessary considering the future inference performance maximization.

To address these challenges, we adopt a two-loop alternating optimization approach. At the outer loop, we alternatively optimize the model training and inference based on the real-time system state to ensure the global model performance is improved. %
Within the outer loop of optimization, we further optimize the download probability and service rate alternatively after clients' participation probability has been decided, referred to as the inner loop of alternating optimization.

We present the online joint federated learning and service provisioning (FedLS) algorithm in Algorithm~\ref{alg:joint}. In FedLS, to avoid a high degree of constraint violation represented by queue lengths, we initialize the queue lengths with a constant $W$ (line \ref{algline:init}). In each time slot $t$, 
each client determines its participation probability $q_t^n$, the download probability  $\beta_{t+1}^n$ and service rate $\mu_{t+1}^n$, using the two-loop alternating optimization.

\textbf{Outer Loop: Alternatively Optimizing Model Training and Inference.}
To ensure the global model performance $G_{t}$ is continuously improved, we assume $\beta_{t+1}^n=1$ to optimize $G_{t+1}$. 
Therefore, in the outer loop, we alternatively optimize the client participation probability $q_t^n$ to improve $G_{t+1}$ (line \ref{algline:get_q}) and the download probability  $\beta_{t+1}^n$ and service rate $\mu_{t+1}^n$ to optimize $M_{t+2}$.
By adopting the same virtual queue drifts and maximum instantaneous resource constraints, the approach effectively balances resources for model training and inference.
 Within the outer loop for fixed $q_t^n$, we leave the optimization of $M_{t+2}$ to the inner loop of alternating optimization.

\textbf{Inner Loop: Alternatively Optimizing Download Probability and Service Rate in Request Processing.}
We can observe that the Problem $\mathbf{P2-}t$ is still non-convex with respective to $\beta_{t+1}^n$ and $\mu_{t+1}^n$ even when $q_t^n$ is known. To address the difficulty, we apply the inner loop of alternating optimization. Namely, we iteratively optimize the download probability  $\beta_{t+1}^n$ and service rate $\mu_{t+1}^n$ till converge to obtain the optimal inference performance at clients (lines \ref{algline:get_bm_start}--\ref{algline:get_bm_end}).

Based on the determined values of $q_t^n$, $\beta_t^n$, and $\mu_t^n$, clients perform request processing and model training in parallel in each time slot. Specifically, each client determines global model download decisions (lines \ref{algline:get_i_1}, \ref{algline:get_i_2}), request processing (line \ref{algline:service_providing}), and model training and aggregation (lines \ref{algline:update_model}, \ref{algline:clientToServer}, \ref{algline:model_aggre}). 
At the end of each time slot, each client updates the service queue $\Lambda_{t}^n$ and virtual queues $\Phi_{t}^n$ and $\Psi_{t}^n$ (line \ref{algline:update_queue}) %
to facilitate the optimization in the subsequent time slot.

\subsection{Analysis of the FedLS Algorithm}\label{sec:constraint_satis}
We discuss the constraint satisfaction for \eqref{eq:cnst_service}--\eqref{eq:cnst_commu} of using Algorithm \ref{alg:joint} to solve Problem $\mathbf{P2-}t$. %
First, some mild assumptions are made as follows.
\begin{assumption} \label{asm:cnst_satis}
    There exist a positive integer $D$ such that $q_t^n \in \left[\frac{1}{D},1\right]$, $G_t \in \left[0,D\right]$, $\tilde{\phi}_n\geq \phi_t^n(\frac{1}{D})$, and $\tilde{\psi}_n\geq \psi_t^n(\frac{1}{D}), \forall n, t$. Moreover, there exist a positive constant $H$ such that $\lambda_t^n$, $\phi_t^n$, $\psi_t^n$, $\mu_t^n$, $\tilde{\phi}^n$, $\tilde{\psi}^n \in [0,\sqrt{2H}], \forall n, t$.
\end{assumption}
In Assumption~\ref{asm:cnst_satis}, we consider that the minimum participation probability of clients is set to $\frac{1}{D}$ for some positive integer $D$. %
Additionally, we assume that the global convergence error $G_t$ is bounded by some positive constant $D$. %
Furthermore, we assume that all queues have non-negative and finite arrival and service rates, which are bounded by some positive constant $H$. Additionally, we assume that the average resource capacities are sufficiently high to accommodate clients with the minimum participation probability $\frac{1}{D}$. 
We have the following theorem for constraint satisfaction.
 \begin{theorem} \label{thm:cnst_satisfaction} 
 Under Assumption~\ref{asm:cnst_satis}, solving P2 for each time $t$ ensures the $O\left(\sqrt{\frac{1}{T^2}+\frac{1}{T}} - \frac{1}{T}\right)$ bounds on constraint violation of \eqref{eq:cnst_service}--\eqref{eq:cnst_commu}.
 \end{theorem}
 The proof of this theorem is similar to \cite{wang2022federated} and we omit the details due to the space limit.  We can observe from the theorem that when the number of running rounds goes to infinite, i.e., $T\rightarrow \infty$, the constraint violation bound goes to zero.

\subsection{Closed-form Solutions for Linear Resource Costs} \label{sec:close-form}
In this part, we consider both model training and inference incur linear computation and communication costs and derive the closed-form solutions in the alternating optimization for solving Problem $\mathbf{P2-}t$. The closed-form solutions make the alternating optimization computationally efficient.
For non-linear resource costs, we can still obtain the decisions by linear search for each variable, during the alternating optimization.

\subsubsection{{Linear Computation and Communication Costs}}
We define the computation cost of client $n$ in time slot $t$ as follows.
\begin{equation}\label{eq:comp_cost}
    \phi_t^n\left(q_t^n, \mu_{t}^n\right):=\alpha\left(\tau  B \xi q_t^n+\mu_{t}^n\right), 
\end{equation}
where $ B$ represents the training batch size and $\alpha$ denotes the unit computation resource cost. Moreover, in the request processing, a client proceeds model inference for the requested data without performing backpropagation for model parameter update. Thus, we use $\xi$ to capture the relative model training cost compared to the model inference cost on a single data.

For communication cost, we denote $\gamma$ as the unit communication cost. The download communication cost of client $n$ in time slot $t$ is formulated as a function of  $q_t^n$ and $\beta_{t}^n$: 
\begin{equation}\label{eq:comm_cost}
 \psi_t^n\left(q_t^n,\beta_{t}^n\right):=\gamma \beta_{t}^n \geq \gamma q_{t}^n.
\end{equation}

\begin{algorithm}[t]
\caption{FedLS: Online joint federated learning and service provisioning algorithm} 
\label{alg:joint} 

{\footnotesize

\KwIn{$\eta,\tau, V, W, \mathcal{D}_n, \forall n$}

\SetKwFor{EachClient}{each client $n = 1, \ldots, N$:}{}{}
\SetKwFor{TheServer}{the server:}{}{}

Initialize $\mathbf{x}_{0}, \Lambda_0^n\leftarrow W, \Phi_0^n \leftarrow W, \Psi_0^n\leftarrow W, \forall n$; \label{algline:init}

\For{$t = 0,\ldots, T-1$}
{
    \EachClient{}{\label{algline:start_train}

        Obtain $q_t^n$ by  by solving Problem $\mathbf{P2-}t$ under $\beta_{t+1}^n=1$; %
        \label{algline:get_q}

        $j\leftarrow 0$, initialize $\beta_{t+1}^n$ and $\mu_{t+1}^n$;
        
        \While{$j \leq j_{\max}$ and $\beta_{t+1}^n$ does not converge \label{algline:get_bm_start}}{
            Obtain $\beta_{t+1}^n$ by solving Problem $\mathbf{P2-}t$; %

            Obtain $\mu_{t+1}^n$ by solving Problem $\mathbf{P2-}t$;%

            $j \leftarrow j+1$; \label{algline:get_bm_end}
            }

        \vspace{3pt}
        \nonl \textbf{Global model download:}
        
        Sample $\mathbf{J}_t^{n} \sim {\rm Bernoulli}\left(\beta_t^n\right)$; \label{algline:get_i_1}

         Sample $\mathbf{I}_t^n \sim {\rm Bernoulli}\left(q_t^n\right)$; \label{algline:get_i_2}
         
        \If{$\mathbf{J}_t^{n}=1$ or $\mathbf{I}_t^n=1$}{
        Download $\mathbf{x}_{t}$ from server; \label{algline:download1}
        }
        
          \vspace{3pt}
        \nonl \textbf{Request processing:}
        
        Process $\mu_t^n$ inference requests of the client; \label{algline:service_providing}

        \vspace{3pt}
        \nonl \textbf{Local training:}

        \If{$\mathbf{I}_t^n=1$}{
    
        $\mathbf{x}_{t,0}^n \leftarrow \mathbf{x}_{t}$; 
        
        \For{$i=0,\ldots,\tau-1$}
        {$\mathbf{x}_{t,i+1}^n \leftarrow \mathbf{x}_{t,i}^n - \frac{\eta}{q_t^n} \mathbf{g}_n\left(\mathbf{x}_{t,i}^n\right)$;\label{algline:update_model}}
        
        Send $\mathbf{x}_{t,\tau}^n$ to the server;  \label{algline:clientToServer}

        }
         \vspace{3pt}
          Update $\Lambda_{t}^n$, $\Phi_{t}^n$, and $\Psi_{t}^n$ by \eqref{eq:queue_update}--\eqref{eq:comm_update};\label{algline:update_queue}      
    }
    
    \TheServer{}{
        $\mathbf{x}_{t+1} \leftarrow \frac{1}{N}\sum_{n=1}^N  \mathbf{x}_{t,\tau}^n$;\label{algline:model_aggre}
    }
}
}
\end{algorithm}

\subsubsection{{Closed-Form Solutions}}
We derive the closed-form solutions for Problem $\mathbf{P2-}t$ with linear resource costs, referred to as $\mathbf{P2-}t\mathbf{-linear}$, in this part.
For the resource cost defined in \eqref{eq:comp_cost} and \eqref{eq:comm_cost}, we have the following proposition. %
\begin{proposition}
\label{pro:1}
    Problem $\mathbf{P2-}t\mathbf{-linear}$ is biconvex. %
\end{proposition}

\noindent{\textbf{Proof of Proposition \ref{pro:1}:} The second-order derivative of \eqref{eq:obj_p2} over $q_t^n$ is positive because the coefficient $C$ is positive. The constraints in Problem $\mathbf{P2-}t\mathbf{-linear}$ are linear to $q_t^n$. Thus, Problem $\mathbf{P2-}t\mathbf{-linear}$ is convex to $q_t^n$. 
Moreover, the objective and constraints in Problem $\mathbf{P2-}t\mathbf{-linear}$ are linear with respect to both $\beta_{t+1}^n$ and $\mu_{t+1}^n$, indicating that Problem $\mathbf{P2-}t\mathbf{-linear}$  is convex in terms of these variables. Therefore, Problem $\mathbf{P2-}t\mathbf{-linear}$ is  biconvex. \qed 

Leveraging the biconvexity of Problem $\mathbf{P2-}t\mathbf{-linear}$, we can employ the alternate convex search approach to find a local optimal solution \cite{gorski2007biconvex, 9705079}. In the following, we present the solutions $\hat{q}_t^n, \hat{\beta}_{t+1}^n$, and $\hat{\mu}_{t+1}^n$ for each variable  ${q}_t^n, {\beta}_{t+1}^n$, and ${\mu}_{t+1}^n$. We omit the detail derivation process due to space limits.

\textbf{Solution to $q_t^n$.}
The solution to $q_t^n$ is equivalent to 
\begin{subequations}
\begin{align}
    \hat{q}_t^n:= &\arg\min_{q_t^n} \frac{VC}{N(t+1)q_t^n}\!+\!\Phi_t^n\alpha\tau  B \xi q_t^n\!+\!\Psi_t^n\gamma q_t^n, \tag{14}\label{eq:p3.1_trans1}\\
    &{\rm s.t.} \quad \alpha\left(\tau  B \xi q_t^n+\hat{\mu}_{t}^n\right) \leq \tilde{\phi}^n_{\max},\label{eq:p31-1}\\
    & \quad\quad \ \gamma q_t^n\leq \tilde{\psi}^n_{\max}.\label{eq:p31-2}
\end{align}
\end{subequations}
It can be seen that the above optimization problem is convex.
By letting the derivative of \eqref{eq:p3.1_trans1} equal to zero, taking into account the boundary constraints in \eqref{eq:p31-1} and \eqref{eq:p31-2}, and noting that $q_t^n\in\left(0,1\right]$, we have
\begin{align} \label{eq:q_solution}
    \hat{q}_t^n\!:=\!\min\left\{1,\sqrt{\frac{VC/N(t+1)}{\Phi_t^n\alpha\tau  B\xi\!+\!\Psi_t^n\gamma}}, \frac{\tilde{\phi}^n_{\max}}{\alpha \tau  B \xi}\!-\!\frac{\hat{\mu}_{t}^n}{ \tau  B \xi},  \frac{\tilde{\psi}^n_{\max}}{\gamma}\right\}.
\end{align}

\textbf{Solution to  $\beta_{t+1}^n$ and  $\mu_{t+1}^n$.} 
For given $\hat{\mu}_{t+1}^n$, the optimal solution to $\beta_{t+1}^n$ is
\begin{align} \label{eq:beta_solution} 
    \hat{\beta}_{t+1}^n\!=\!\left\{\begin{aligned}
        &0, {\rm if}\ V\hat{\mu}_{t+1}^n(G_{t+1}\!-\!G_{t}\hat{q}_t^n\!-\!K_{t}^n(1\!-\!\hat{q}_t^n))\!+\!\Psi_t^n\gamma\!>\!0,\\
        &\textstyle\min\left\{1,\sfrac{\tilde{\psi}^n_{\max}}{\gamma}\right\}, {\rm otherwise.}
    \end{aligned}\right.
\end{align}

On the other hand, for fixed $\hat{\beta}_{t+1}^n$ and noting that $\mu_{t+1}^n\in\left[0,\Lambda_t^n\right]$, the optimal solution of $\mu_{t+1}^n$ is given by
\begin{align} \label{eq:mu_solution}
    \hat{\mu}_{t+1}^n=\left\{\begin{aligned}
        &0, {\rm if}\ V\left(G_{t+1}\hat{\beta}^n_{t+1}\!+G_{t}\left(1-\hat{\beta}_{t+1}^n\right)\hat{q}_{t}^n\right.\\
    & \left.+K_{t}^n\left(1-\!\hat{\beta}_{t+1}^n\!\right)\!\left(1\!-\!\hat{q}_{t}^n\right)\right)\!-\!\Lambda_t^n\!+\!\Phi_t^n\alpha>0,\\
        &\textstyle\min\!\left\{\Lambda_t^n,\frac{\tilde{\phi}_{\max}^n}{\alpha}\!-\!\tau  B \xi \hat{q}_t^n\!\right\}, {\rm otherwise.}
    \end{aligned}\right.
\end{align}
In the equation, since we cannot know the exact value of $q_{t+1}^n$  in the current time slot $t$ in advance, we employ $\hat{q}_{t}^n$ instead in the maximum computation constraint to obtain the boundary value of $\hat{\mu}_{t+1}^n$. In the subsequent time  slot, we will constrain $\hat{q}_{t+1}^n$ based on $\hat{\mu}_{t+1}^n$ as shown in \eqref{eq:q_solution}. By doing so, we ensure that the computation cost is always constrained by $\tilde{\phi}^n_{\max}$.

\begin{figure}[t]
\centering
 \includegraphics[width=0.443\textwidth]{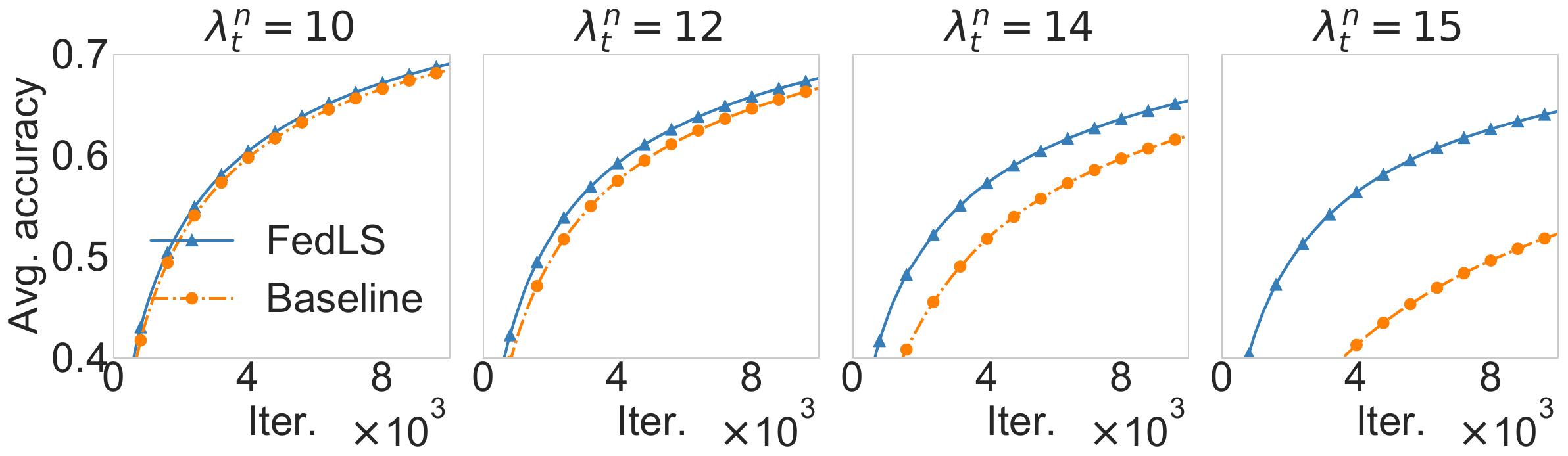}
 \includegraphics[width=0.441\textwidth]{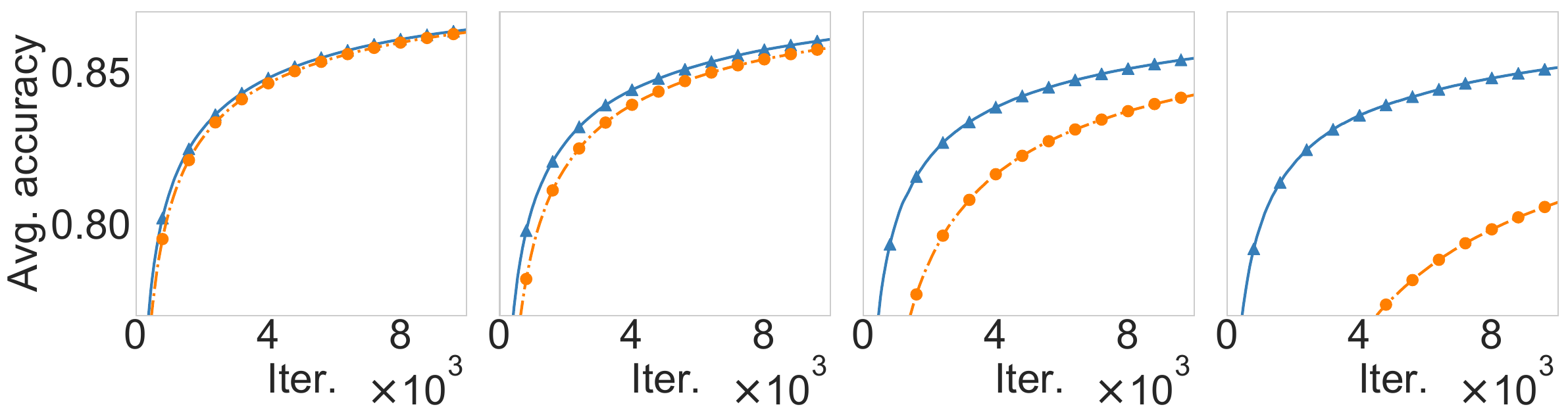}
 \includegraphics[width=0.441\textwidth]{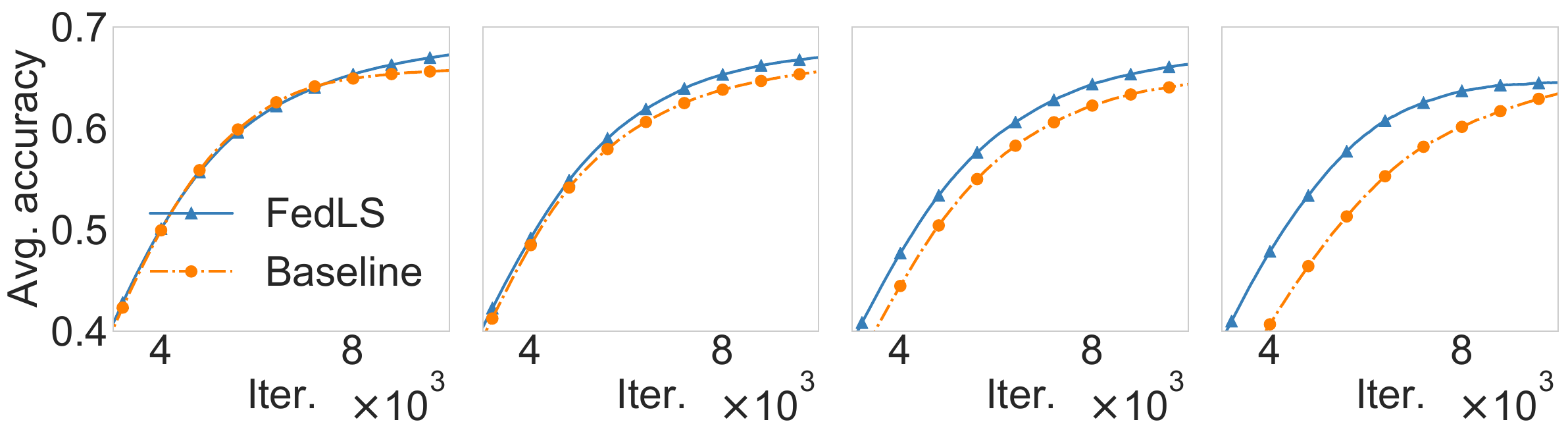}
\caption{\label{fig:perf-lambda}  Inference performance under different request arrival rates, from top to bottom: CIFAR-10, Fashion-MNIST, and SVHN.}
\end{figure}

\begin{figure}[t]
\centering
 \includegraphics[width=0.443\textwidth]{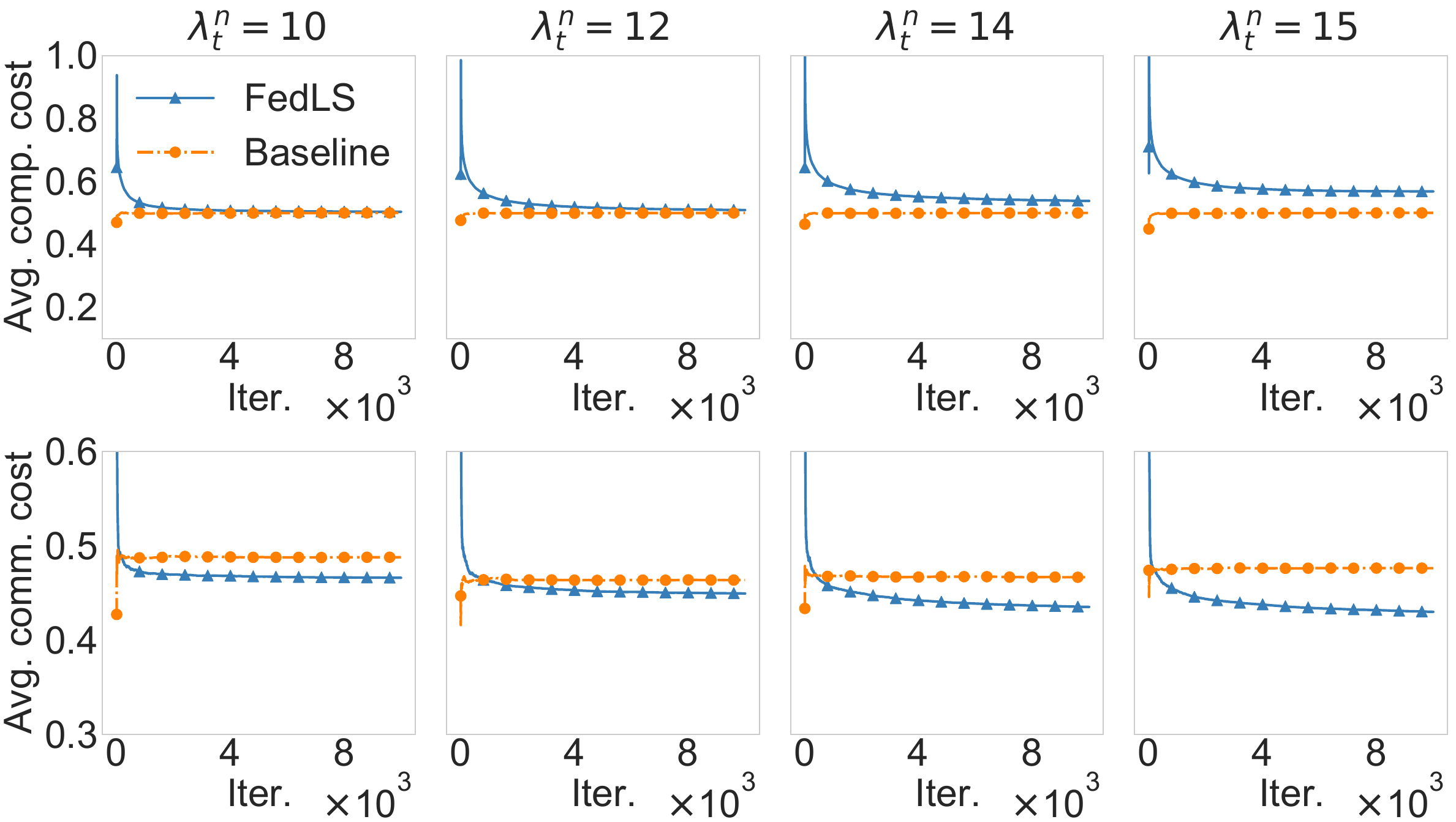}
\caption{\label{fig:cifar-cost}  Resource costs of CIFAR-10 under different request arrival rates.}
\end{figure}

\textbf{Insights on clients' request processing behavior.} We can observe from \eqref{eq:mu_solution} that the service rate $\hat{\mu}_t^n$ can only take three possible values:  the current service queue length, the maximum allowable service rate determined by the computation resource constraint, or zero.  %
A client tends to serve requests at a later time to obtain a high inference performance.  However, when considering the mean rate state constraint in \eqref{eq:cnst_service}, the client chooses to serve a subset of requests simultaneously. This approach helps the client avoid a high degree of constraint violation while still ensuring that each request receives a high inference performance. 

\section{Experiments}
\label{sec:experimentation}
In the experiments, we conduct experiments on real-world datasets to assess the performance of the FedLS. Specifically, we first evaluate the performance of FedLS in inference accuracy and processing delay. Then, we evaluate the robustness of FedLS for varying coefficients $V$ and $W$. %
We implement all models with PyTorch, and all experiments are on a server with RTX 3090 GPUs.

\subsection{Setup}
\subsubsection{Datasets and Hyper-parameters}
We conduct experiments on Fashion-MNIST \cite{xiao2017fashion}, SVHN \cite{netzer2011reading}, and CIFAR-10 \cite{krizhevsky2009learning} datasets for 100 clients. We simulate a high data heterogeneity setup by partitioning the datasets in a non-i.i.d. manner so that each client has data of one class for Fashion-MNIST and CIFAR-10 and two classes for SVHN. 
We use convolutional neural network (CNN) models %
to train the datasets and set the learning rate as $\eta=0.1$ for Fashion-MNIST and CIFAR-10 and  $\eta=0.05$ for SVHN. We apply stochastic gradient descent (SGD) optimizer and set $\tau=1$ and $ B=16$. 

We evaluate the linear computation and communication costs and set the average and maximum resource constraints as $\tilde{\phi}^n=0.5$, $\tilde{\phi}^n_{\max}=5$, $\tilde{\psi}^n=0.5$, and $\tilde{\psi}^n_{\max}=5$.
We randomly generate the time-varying instantaneous resource cost following \cite{wang2022federated}, where computation cost coefficient $\alpha$ follows a uniform random distribution with the mean of 0.03 and the communication cost coefficient $\gamma$ is related to channel capacity with Rayleigh fading. We set the coefficient $\xi=2$. 

We determine the constant parameter $C$ in convergence error by evaluating the gradient norm along with model training and find $C=10^{-6}$. The arrival rates of inference requests on clients follow Poisson distribution. We will evaluate the effect of different request arrival rates in our experiments.
For Lyanouov optimization, we set $V=1$ and $W=1$ by default unless specifically mentioned.

\subsubsection{Baseline}
Since there is not any work considering the resource balancing between model training and inference, we consider a baseline algorithm by setting the variables according to the average available resources. Specifically, we first set the service rate equal to the arrival rate to ensure queue stability. Then, we determine the participation probability considering the average computation and communication resource budgets\footnote{Generally, the instantaneous maximum allowable resource cost is higher than the average resource cost. Thus, we only consider average resource cost as budget in the baseline.} as
    $q_t^n=\min\left\{1,\frac{\tilde{\phi}^n-\alpha\lambda_t^n}{\alpha\tau B\xi},\frac{\tilde{\psi}^n}{\gamma}\right\}, \forall t, n.$
Similarly, we set the download probabilities and service rates as $\beta_{t}^n=\min\left\{1, \frac{\tilde{\psi}^n}{\gamma}\right\}$ and $\mu_{t}^n=\min\left\{\Lambda_t^n,\frac{\tilde{\phi}^n}{\alpha}-\tau  B \xi q_t^n\right\}, \forall t, n$. The construction of this baseline is based on similar ideas as that in \cite{wang2022federated}, although \cite{wang2022federated} solves a different problem without considering joint model training and inference.

\begin{figure}[t]
\centering
\includegraphics[width=0.48\textwidth]{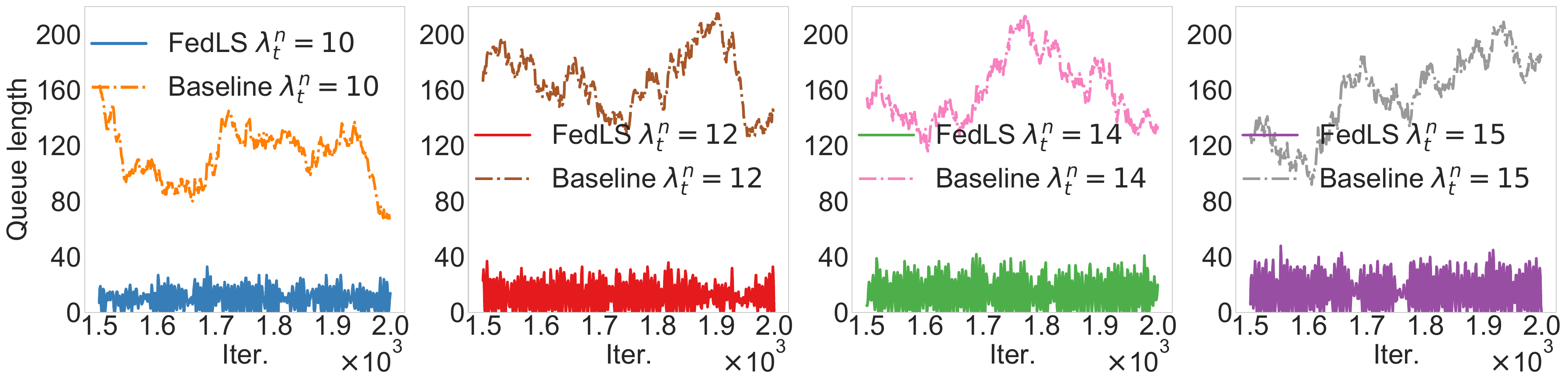}
\caption{\label{fig:cifar-lambda-queue}  Queue lengths of CIFAR-10 under different request arrival rates.}
\end{figure}

\begin{figure}[t]
\centering
\begin{subfigure}[b]{0.125\columnwidth}
\centering
\includegraphics[width=1\textwidth]{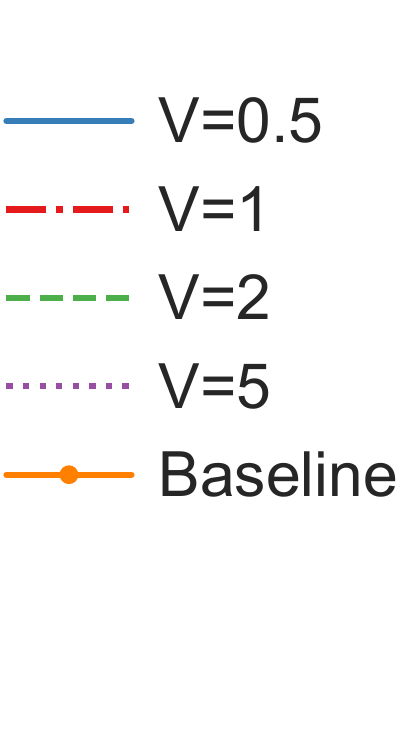}
\end{subfigure}
\begin{subfigure}[b]{0.25\columnwidth}
 \centering
 \includegraphics[width=\textwidth]{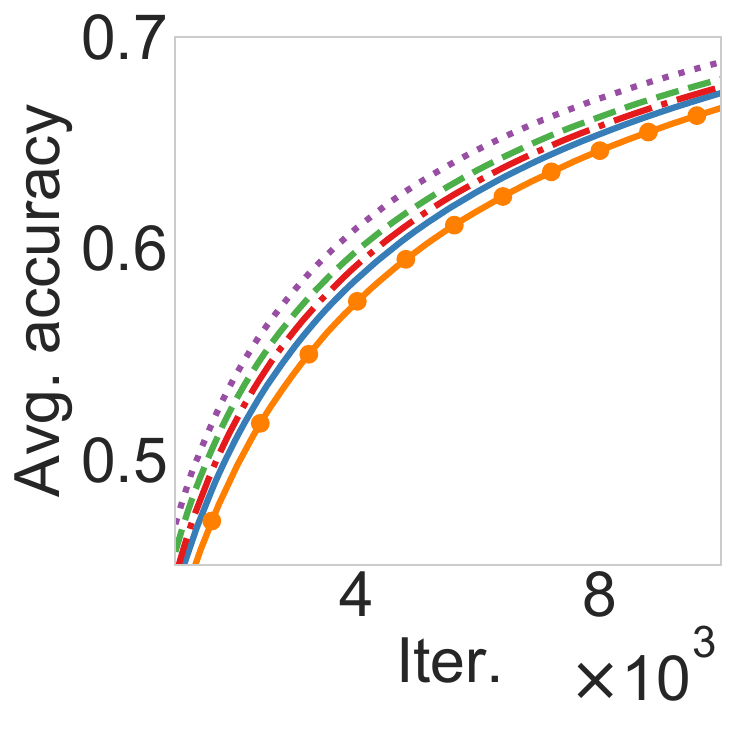}
\end{subfigure}
\begin{subfigure}[b]{0.25\columnwidth}
 \centering
 \includegraphics[width=\textwidth]{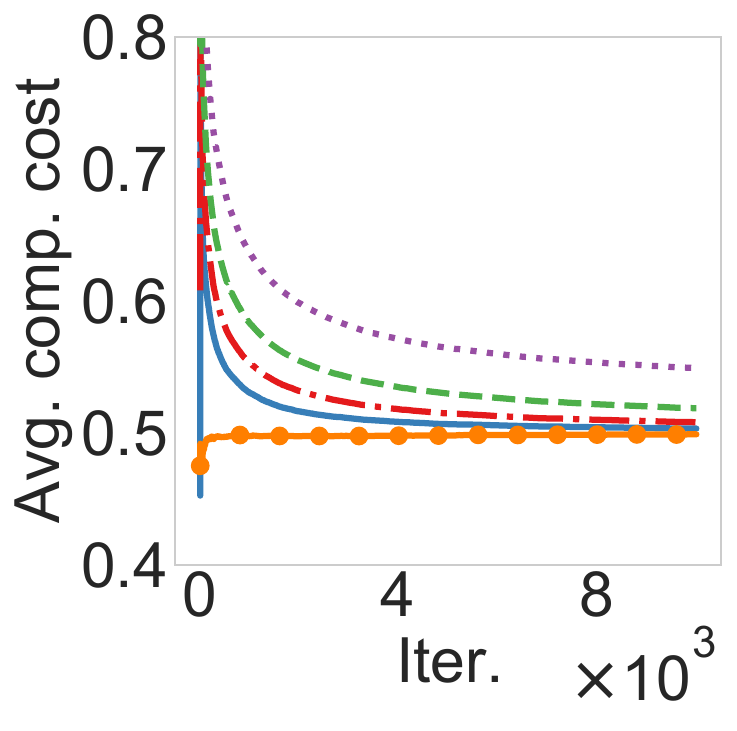}
\end{subfigure}
\begin{subfigure}[b]{0.25\columnwidth}
 \centering
 \includegraphics[width=\textwidth]{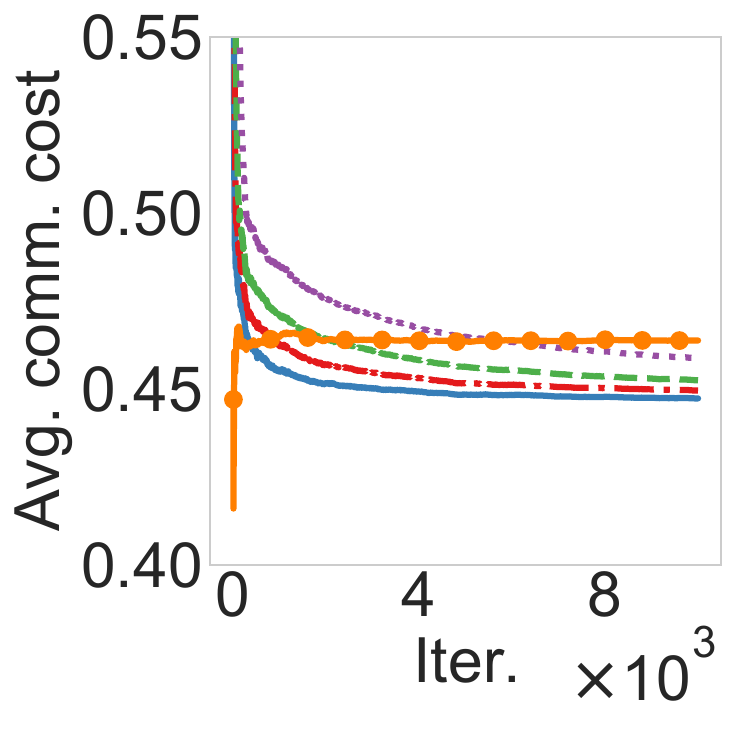}
\end{subfigure}
\caption{\label{fig:cifar-v}  Performance of CIFAR-10 under different coefficient $V$.}
\vspace{5pt}
\centering
\begin{subfigure}[b]{0.125\columnwidth}
\centering
\includegraphics[width=1\textwidth]{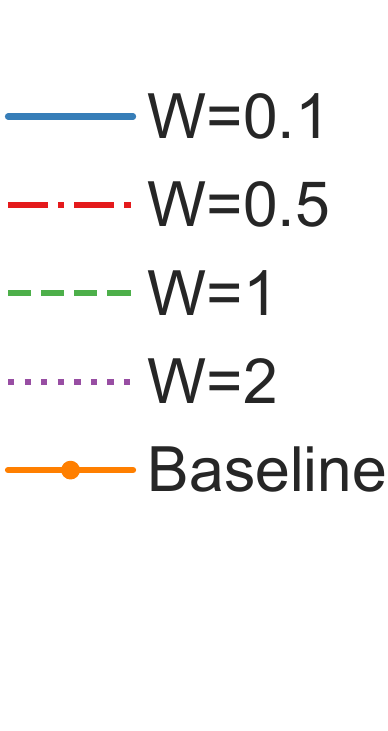}
\end{subfigure}
\begin{subfigure}[b]{0.25\columnwidth}
 \centering
 \includegraphics[width=\textwidth]{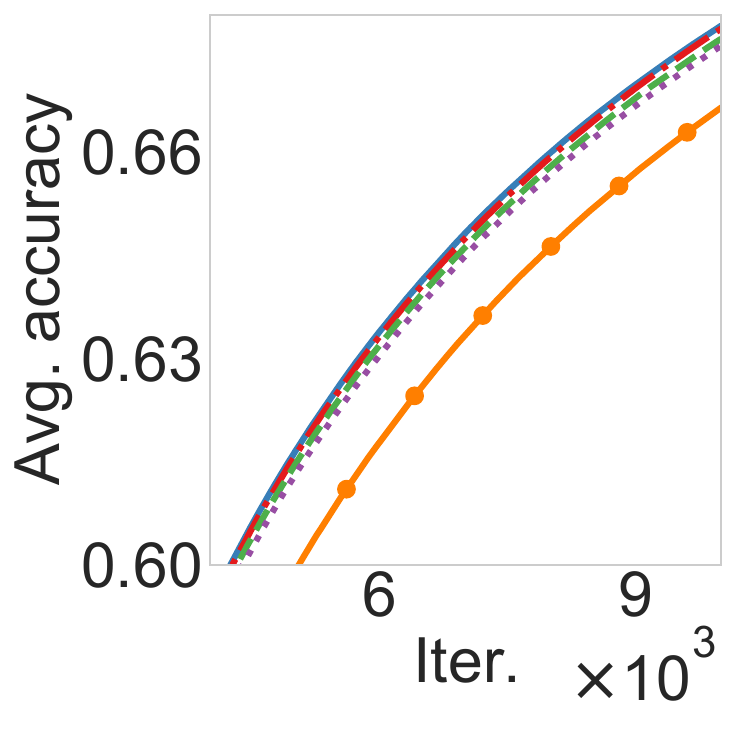}
\end{subfigure}
\begin{subfigure}[b]{0.25\columnwidth}
 \centering
 \includegraphics[width=\textwidth]{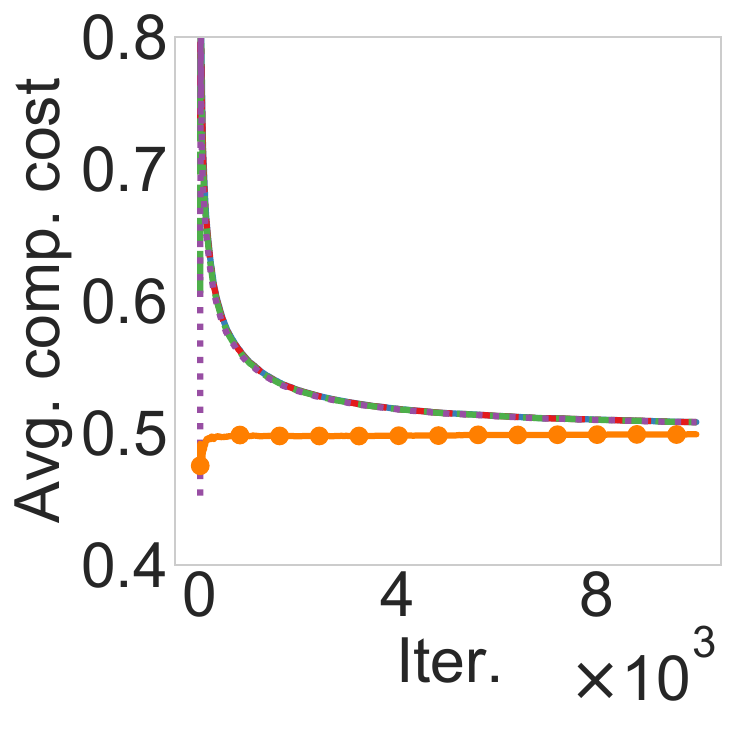}
\end{subfigure}
\begin{subfigure}[b]{0.25\columnwidth}
 \centering
 \includegraphics[width=\textwidth]{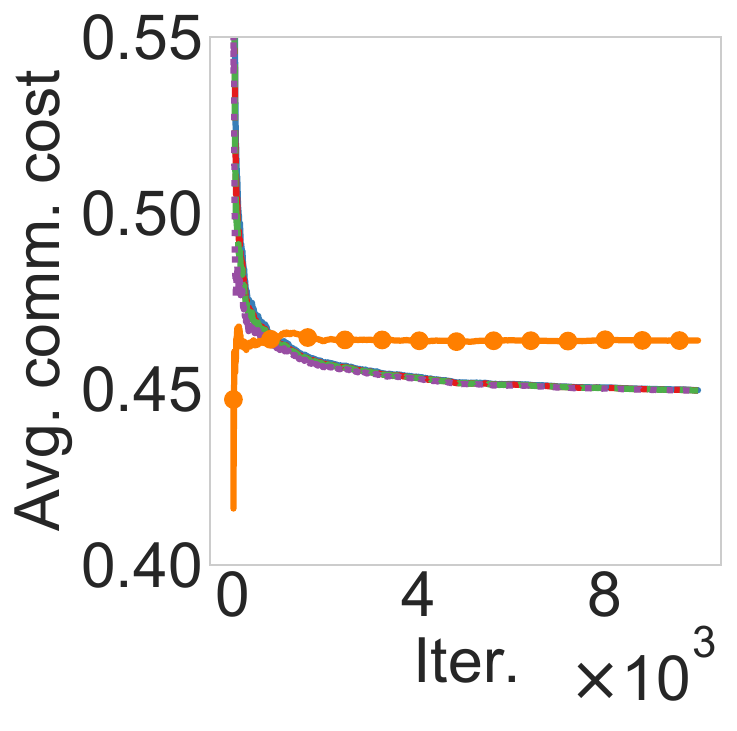}
\end{subfigure}
\caption{\label{fig:cifar-w}  Performance of CIFAR-10 under different coefficient $W$.}
\end{figure}

\subsection{Results}
\subsubsection{{Performance Over Varying Request Arrival Rates}}
We evaluate the performance of FedLS in inference accuracy, resource costs, and processing delay.

\textbf{Performance of Inference Accuracy and Resource Costs.} We evaluate the performance of the FedLS algorithm against the baseline by comparing their average inference accuracy and average computation and communication costs. Figure~\ref{fig:perf-lambda} shows the results for different datasets under varying request arrival rates ($\lambda_t^n$), with fixed coefficients $V=1$ and $W=1$. Our experiments demonstrate that FedLS consistently outperforms the baseline in the inference accuracy at clients.
As expected, increasing the inference request arrival rate (as indicated by higher $\lambda_t^n$) negatively impacts the average inference performance at clients. This is because processing more requests consumes more resources, leaving less resources available for model training. We observe that the advantage of FedLS over the baseline becomes more remarkable at higher request arrival rates. For instance, at $\lambda_t^n=15$, the inference accuracy of the FedLS is 12\% better than that of the baseline for CIFAR-10.

The average computation and communication costs approach or are below the resource limitations as time goes longer, indicating that the time-average resource constraints are satisfied, as shown in Fig. \ref{fig:cifar-cost}.
The resource cost results of Fashion-MNIST and SVHN show the same observation and conclusion as CIFAR-10. 

\textbf{Performance of Processing Delay.} 
Figure \ref{fig:cifar-lambda-queue} provides examples of instantaneous service queue lengths for an individual client, which clearly demonstrate the superiority of FedLS in low processing delay. We observe from Fig. \ref{fig:cifar-lambda-queue} that both FedLS and the baseline ensure the mean rate stability of the service queue because the queue lengths do not go infinite along with time.  However, the processing delay of the baseline is significantly higher than that of our FedLS for all inference request arrival rates. To see this, 
while the queue lengths for FedLS increase with higher request arrival rates, they remain below 40. In contrast, the baseline method may experience queue lengths exceeding 100 under the same scenarios.

\subsubsection{{Performance under Different Coefficients}}
We evaluate the robustness of FedLS for varying coefficients of $V$ and $W$ for CIFAR-10 dataset with $\lambda_t^n=12$. Figure \ref{fig:cifar-v} demonstrates the performance of FedLS for fixed $W=1$ and varying $V=0.5, 1, 2,$ and 5 and Fig. \ref{fig:cifar-w} shows the performance of FedLS for fixed $V=1$ and varying $W=0.1, 0.5, 1,$ and 2. We observe that FedLS consistently obtains better performance than the baseline for different coefficients.

\section{Conclusion}\label{sec:conclusion} 
In this paper, we highlighted the periodic model upgrade using newly collected data for online applications. We focused on the coexistence of FL and SP to optimize the inference performance for request processing with service stability, under the computation and communication resource constraints. First, we address the challenge of characterizing the inference performance by analyzing the convergence error of FL and introducing the notion of AoM to capture the client-side model freshness. Then, we approximate the long-term inference performance optimization problem using an online decision-making framework and proposed the FedLS algorithm to adaptively balance the resource for model training and inference. Experimental results demonstrate the effectiveness of our algorithm in improving the averaging inference accuracy with low processing delay.

\clearpage

\bibliographystyle{IEEEtran}
\bibliography{ref}

\begin{thebibliography}{10}
\providecommand{\url}[1]{#1}
\csname url@samestyle\endcsname
\providecommand{\newblock}{\relax}
\providecommand{\bibinfo}[2]{#2}
\providecommand{\BIBentrySTDinterwordspacing}{\spaceskip=0pt\relax}
\providecommand{\BIBentryALTinterwordstretchfactor}{4}
\providecommand{\BIBentryALTinterwordspacing}{\spaceskip=\fontdimen2\font plus
\BIBentryALTinterwordstretchfactor\fontdimen3\font minus
  \fontdimen4\font\relax}
\providecommand{\BIBforeignlanguage}[2]{{%
\expandafter\ifx\csname l@#1\endcsname\relax
\typeout{** WARNING: IEEEtran.bst: No hyphenation pattern has been}%
\typeout{** loaded for the language `#1'. Using the pattern for}%
\typeout{** the default language instead.}%
\else
\language=\csname l@#1\endcsname
\fi
#2}}
\providecommand{\BIBdecl}{\relax}
\BIBdecl

\bibitem{TorchServe}
``Torchserve,'' \url{https://pytorch.org/serve}.

\bibitem{TensorFlowServing}
``Tensorflow serving,'' \url{https://github.com/tensorflow/serving}.

\bibitem{mcmahan2017communication}
B.~McMahan, E.~Moore, D.~Ramage, S.~Hampson, and B.~A. y~Arcas,
  ``Communication-efficient learning of deep networks from decentralized
  data,'' in \emph{Artificial intelligence and statistics}.\hskip 1em plus
  0.5em minus 0.4em\relax PMLR, 2017, pp. 1273--1282.

\bibitem{kairouz2019advances}
P.~Kairouz, H.~B. McMahan, B.~Avent, A.~Bellet, M.~Bennis, A.~N. Bhagoji,
  K.~Bonawitz, Z.~Charles, G.~Cormode, R.~Cummings \emph{et~al.}, ``Advances
  and open problems in federated learning,'' \emph{Foundations and
  Trends{\textregistered} in Machine Learning}, vol.~14, no. 1--2, pp. 1--210,
  2021.

\bibitem{sun2021pain}
P.~Sun, H.~Che, Z.~Wang, Y.~Wang, T.~Wang, L.~Wu, and H.~Shao, ``Pain-fl:
  Personalized privacy-preserving incentive for federated learning,''
  \emph{IEEE Journal on Selected Areas in Communications}, vol.~39, no.~12, pp.
  3805--3820, 2021.

\bibitem{sun2022profit}
P.~Sun, X.~Chen, G.~Liao, and J.~Huang, ``A profit-maximizing model marketplace
  with differentially private federated learning,'' in \emph{IEEE INFOCOM
  2022-IEEE Conference on Computer Communications}.\hskip 1em plus 0.5em minus
  0.4em\relax IEEE, 2022, pp. 1439--1448.

\bibitem{imteaj2021survey}
A.~Imteaj, U.~Thakker, S.~Wang, J.~Li, and M.~H. Amini, ``A survey on federated
  learning for resource-constrained iot devices,'' \emph{IEEE Internet of
  Things Journal}, vol.~9, no.~1, pp. 1--24, 2021.

\bibitem{Cho2022towards}
Y.~J. Cho, J.~Wang, and G.~Joshi, ``Towards understanding biased client
  selection in federated learning,'' in \emph{International Conference on
  Artificial Intelligence and Statistics}.\hskip 1em plus 0.5em minus
  0.4em\relax PMLR, 2022, pp. 10\,351--10\,375.

\bibitem{zhou2023rein}
R.~Zhou, J.~Yu, R.~Wang, B.~Li, J.~Jiang, and L.~Wu, ``A reinforcement learning
  approach for minimizing job completion time in clustered federated
  learning,'' in \emph{IEEE INFOCOM}, 2023.

\bibitem{ding2023incentive}
L.~G. Ningning~Ding and J.~Huang, ``Joint participation incentive and network
  pricing design for federated learning,'' in \emph{IEEE INFOCOM}, 2023.

\bibitem{jiang2023hetero}
Z.~Jiang, Y.~Xu, H.~Xu, Z.~Wang, and C.~Qian, ``Heterogeneity-aware federated
  learning with adaptive client selection and gradient compression,'' in
  \emph{IEEE INFOCOM}, 2023.

\bibitem{9484767}
J.~Liu, H.~Xu, L.~Wang, Y.~Xu, C.~Qian, J.~Huang, and H.~Huang, ``Adaptive
  asynchronous federated learning in resource-constrained edge computing,''
  \emph{IEEE Transactions on Mobile Computing}, vol.~22, no.~2, pp. 674--690,
  2023.

\bibitem{9435350}
M.~Salehi and E.~Hossain, ``Federated learning in unreliable and
  resource-constrained cellular wireless networks,'' \emph{IEEE Transactions on
  Communications}, vol.~69, no.~8, pp. 5136--5151, 2021.

\bibitem{9302575}
R.~Saha, S.~Misra, and P.~K. Deb, ``Fogfl: Fog-assisted federated learning for
  resource-constrained iot devices,'' \emph{IEEE Internet of Things Journal},
  vol.~8, no.~10, pp. 8456--8463, 2021.

\bibitem{fraboni2021clustered}
Y.~Fraboni, R.~Vidal, L.~Kameni, and M.~Lorenzi, ``Clustered sampling:
  Low-variance and improved representativity for clients selection in federated
  learning,'' in \emph{ICML}.\hskip 1em plus 0.5em minus 0.4em\relax PMLR,
  2021, pp. 3407--3416.

\bibitem{yang2021achieving}
H.~Yang, M.~Fang, and J.~Liu, ``Achieving linear speedup with partial worker
  participation in non-iid federated learning,'' in \emph{ICLR}, 2021.

\bibitem{li2019convergence}
X.~Li, K.~Huang, W.~Yang, S.~Wang, and Z.~Zhang, ``On the convergence of fedavg
  on non-iid data,'' in \emph{ICLR}, 2020.

\bibitem{9053740}
H.~H. Yang, A.~Arafa, T.~Q.~S. Quek, and H.~Vincent~Poor, ``Age-based
  scheduling policy for federated learning in mobile edge networks,'' in
  \emph{IEEE ICASSP}, 2020, pp. 8743--8747.

\bibitem{wang2022age}
K.~Wang, Y.~Ma, M.~B. Mashhadi, C.~H. Foh, R.~Tafazolli, and Z.~Ding, ``Age of
  information in federated learning over wireless networks,'' \emph{arXiv
  preprint arXiv:2209.06623}, 2022.

\bibitem{maaoi}
M.~Ma, V.~W. Wong, and R.~Schober, ``Aoi-driven client scheduling for federated
  learning: A lagrangian index approach.''

\bibitem{wang2019adaptive}
S.~Wang, T.~Tuor, T.~Salonidis, K.~K. Leung, C.~Makaya, T.~He, and K.~Chan,
  ``Adaptive federated learning in resource constrained edge computing
  systems,'' \emph{IEEE Journal on Selected Areas in Communications}, vol.~37,
  no.~6, pp. 1205--1221, 2019.

\bibitem{liao2023adap}
Y.~Liao, Y.~Xu, H.~Xu, L.~Wang, and C.~Qian, ``Adaptive configuration for
  heterogeneous participants in decentralized federated learning,'' in
  \emph{IEEE INFOCOM}, 2023.

\bibitem{8889996}
F.~Sattler, S.~Wiedemann, K.-R. Müller, and W.~Samek, ``Robust and
  communication-efficient federated learning from non-i.i.d. data,'' \emph{IEEE
  Transactions on Neural Networks and Learning Systems}, vol.~31, no.~9, pp.
  3400--3413, 2020.

\bibitem{gorbunov2021marina}
E.~Gorbunov, K.~P. Burlachenko, Z.~Li, and P.~Richt{\'a}rik, ``Marina: Faster
  non-convex distributed learning with compression,'' in \emph{ICML}.\hskip 1em
  plus 0.5em minus 0.4em\relax PMLR, 2021, pp. 3788--3798.

\bibitem{alistarh2018convergence}
D.~Alistarh, T.~Hoefler, M.~Johansson, N.~Konstantinov, S.~Khirirat, and
  C.~Renggli, ``The convergence of sparsified gradient methods,''
  \emph{Advances in Neural Information Processing Systems}, vol.~31, 2018.

\bibitem{stich2019error}
S.~U. Stich and S.~P. Karimireddy, ``The error-feedback framework: Better rates
  for sgd with delayed gradients and compressed communication,'' \emph{Journal
  of Machine Learning Research}, vol.~21, no. 237, p. 1–36, 2020.

\bibitem{hegde2023network}
A.~M. Parikshit~Hegde, Gustavo de~Veciana, ``Network adaptive federated
  learning: Congestion and lossy compression,'' in \emph{INFOCOM}, 2023.

\bibitem{li2023anyc}
P.~Li, G.~Cheng, X.~Huang, J.~Kang, R.~Yu, Y.~Wu, and M.~Pan, ``Anycostfl:
  Efficient on-demand federated learning over heterogeneous edge devices,'' in
  \emph{IEEE INFOCOM}, 2023.

\bibitem{9707474_dropout}
D.~Wen, K.-J. Jeon, and K.~Huang, ``Federated dropout—a simple approach for
  enabling federated learning on resource constrained devices,'' \emph{IEEE
  Wireless Communications Letters}, vol.~11, no.~5, pp. 923--927, 2022.

\bibitem{Alistarh2017qsgd}
D.~Alistarh, D.~Grubic, J.~Li, R.~Tomioka, and M.~Vojnovic, ``Qsgd:
  Communication-efficient sgd via gradient quantization and encoding,''
  \emph{Advances in neural information processing systems}, vol.~30, 2017.

\bibitem{reisizadeh2020fedpaq}
A.~Reisizadeh, A.~Mokhtari, H.~Hassani, A.~Jadbabaie, and R.~Pedarsani,
  ``Fedpaq: A communication-efficient federated learning method with periodic
  averaging and quantization,'' in \emph{International Conference on Artificial
  Intelligence and Statistics}.\hskip 1em plus 0.5em minus 0.4em\relax PMLR,
  2020, pp. 2021--2031.

\bibitem{liu2023commu}
F.~H. Heting~Liu and G.~Cao, ``Communication-efficient federated learning for
  heterogeneous edge devices based on adaptive gradient quantization,'' in
  \emph{IEEE INFOCOM}, 2023.

\bibitem{wang2022fedlite}
J.~Wang, H.~Qi, A.~S. Rawat, S.~Reddi, S.~Waghmare, F.~X. Yu, and G.~Joshi,
  ``Fedlite: A scalable approach for federated learning on resource-constrained
  clients,'' \emph{arXiv preprint arXiv:2201.11865}, 2022.

\bibitem{8514816}
I.~Kadota, A.~Sinha, E.~Uysal-Biyikoglu, R.~Singh, and E.~Modiano, ``Scheduling
  policies for minimizing age of information in broadcast wireless networks,''
  \emph{IEEE/ACM Transactions on Networking}, vol.~26, no.~6, pp. 2637--2650,
  2018.

\bibitem{9484640}
M.~K. Chowdhury~Shisher, H.~Qin, L.~Yang, F.~Yan, and Y.~Sun, ``The age of
  correlated features in supervised learning based forecasting,'' in \emph{IEEE
  INFOCOM WKSHPS}, 2021, pp. 1--8.

\bibitem{wang2022federated}
S.~Wang, J.~Perazzone, M.~Ji, and K.~S. Chan, ``Federated learning with
  flexible control,'' in \emph{IEEE INFOCOM}, 2023.

\bibitem{luo2021cost}
B.~Luo, X.~Li, S.~Wang, J.~Huang, and L.~Tassiulas, ``Cost-effective federated
  learning in mobile edge networks,'' \emph{IEEE Journal on Selected Areas in
  Communications}, vol.~39, no.~12, pp. 3606--3621, 2021.

\bibitem{bookneely}
M.~Neely, \emph{Stochastic Network Optimization with Application to
  Communication and Queueing Systems}, 01 2010, vol.~3.

\bibitem{gorski2007biconvex}
J.~Gorski, F.~Pfeuffer, and K.~Klamroth, ``Biconvex sets and optimization with
  biconvex functions: a survey and extensions,'' \emph{Mathematical methods of
  operations research}, vol.~66, pp. 373--407, 2007.

\bibitem{9705079}
Y.~Jiao, K.~Yang, D.~Song, and D.~Tao, ``Timeautoad: Autonomous anomaly
  detection with self-supervised contrastive loss for multivariate time
  series,'' \emph{IEEE Transactions on Network Science and Engineering},
  vol.~9, no.~3, pp. 1604--1619, 2022.

\bibitem{pmlr-v119-karimireddy20a}
S.~P. Karimireddy, S.~Kale, M.~Mohri, S.~Reddi, S.~Stich, and A.~T. Suresh,
  ``{SCAFFOLD}: Stochastic controlled averaging for federated learning,'' in
  \emph{ICML}, ser. Proceedings of Machine Learning Research, H.~D. III and
  A.~Singh, Eds., vol. 119.\hskip 1em plus 0.5em minus 0.4em\relax PMLR, 13--18
  Jul 2020, pp. 5132--5143.

\bibitem{50448}
S.~Reddi, Z.~B. Charles, M.~Zaheer, Z.~Garrett, K.~Rush, J.~Konečný,
  S.~Kumar, and B.~McMahan, Eds., \emph{Adaptive Federated Optimization}, 2021.

\bibitem{jhunjhunwala2023fedexp}
D.~Jhunjhunwala, S.~Wang, and G.~Joshi, ``Fedexp: Speeding up federated
  averaging via extrapolation,'' in \emph{ICLR}, 2023.

\bibitem{yu2019parallel}
H.~Yu, S.~Yang, and S.~Zhu, ``Parallel restarted sgd with faster convergence
  and less communication: Demystifying why model averaging works for deep
  learning,'' in \emph{AAAI}, vol.~33, no.~01, 2019, pp. 5693--5700.

\bibitem{perazzone2022communication}
J.~Perazzone, S.~Wang, M.~Ji, and K.~S. Chan, ``Communication-efficient device
  scheduling for federated learning using stochastic optimization,'' in
  \emph{IEEE INFOCOM}.\hskip 1em plus 0.5em minus 0.4em\relax IEEE, 2022, pp.
  1449--1458.

\bibitem{xiao2017fashion}
H.~Xiao, K.~Rasul, and R.~Vollgraf, ``Fashion-mnist: a novel image dataset for
  benchmarking machine learning algorithms,'' \emph{arXiv preprint
  arXiv:1708.07747}, 2017.

\bibitem{netzer2011reading}
Y.~Netzer, T.~Wang, A.~Coates, A.~Bissacco, B.~Wu, and A.~Y. Ng, ``Reading
  digits in natural images with unsupervised feature learning,'' 2011.

\bibitem{krizhevsky2009learning}
A.~Krizhevsky, G.~Hinton \emph{et~al.}, ``Learning multiple layers of features
  from tiny images,'' 2009.

\end{thebibliography}

\end{document}